\documentclass[twocolumn]{aastex631}
\usepackage{tikz}
\usepackage{graphicx}
\usepackage{tabularx}
\usepackage{xcolor,colortbl}
\usepackage{float}
\usepackage{amsmath}
\usepackage{enumitem}

\newcommand{\K}{\ensuremath{\textrm{ K}}}
\newcommand{\m}{\ensuremath{\textrm{ m}}}

\newcommand{\radmsq}{\ensuremath{\textrm{ rad m}^{-2}}}

\shortauthors{A. Ordog et al.}
\shorttitle{CGPS+GMIMS-HBN full-scale RMs}
\begin{document}

\title{A pioneering experiment combining single-antenna and aperture-synthesis data to measure Faraday rotation with GMIMS and the CGPS}

\correspondingauthor{Anna Ordog}
\email{anna.ordog@ubc.ca}

\newcommand{\UBCO}{Department of Computer Science, Math, Physics, \& Statistics, University of British Columbia, Okanagan Campus, Kelowna, BC V1V 1V7, Canada}

\newcommand{\UCalgary}{Department of Physics and Astronomy, University of Calgary, 2500 University Drive NW, Calgary, Alberta, T2N 1N4, Canada}

\newcommand{\DRAO}{Dominion Radio Astrophysical Observatory, Herzberg Research Centre for Astronomy and Astrophysics, National Research Council Canada, PO Box 248, Penticton, BC V2A 6J9, Canada}

\newcommand{\UTas}{School of Natural Sciences, University of Tasmania, Hobart, Tas 7000 Australia}

\newcommand{\INAF}{INAF-Istituto di Radioastronomia, Via Gobetti 101, 40129 Bologna, Italy}

\newcommand{\INAFOAA}{INAF – Osservatorio Astrofisico di Arcetri, Largo E. Fermi 5, 50125 Firenze, Italy}

\newcommand{\LPENS}{Laboratoire de Physique de l’École Normale Supérieure, ENS, Université PSL, CNRS, Sorbonne Université, Université de Paris, F-75005 Paris, France}

\newcommand{\RU}{Department of Astrophysics/IMAPP, Radboud University, PO Box 9010, 6500 GL Nijmegen, The Netherlands}

\newcommand{\CSIROBentley}{ATNF, CSIRO Space \& Astronomy, Bentley, WA, Australia}

\newcommand{\OC}{Department of Physics and Astronomy, Okanagan College, Kelowna, BC V1Y 4X8, Canada}

\newcommand{\Mayo}{Departments of Radiology and Information Technology, Mayo Clinic, 200 1st. St. SW, Rochester, MN 55905, USA}
\author[0000-0002-2465-8937]{Anna Ordog}
\affiliation{\UBCO }
\affiliation{\DRAO }
\affiliation{\UCalgary }

\author[0000-0003-4781-5701]{Jo-Anne C. Brown}
\affiliation{\UCalgary }

\author[0000-0003-1455-2546]{T.~L. Landecker}
\affiliation{\DRAO }

\author[0000-0001-7301-5666]{Alex S. Hill}
\affiliation{\UBCO }
\affiliation{\DRAO }

\author[0000-0001-5953-0100]{Roland Kothes}
\affiliation{\DRAO }

\author[0000-0001-7722-8458]{Jennifer L. West}
\affiliation{\DRAO }

\author[0000-0002-6300-7459]{John M. Dickey}
\affiliation{\UTas }

\author[0000-0002-5288-312X]{Marijke Haverkorn}
\affiliation{\RU }

\author[0000-0002-3973-8403]{Ettore Carretti}
\affiliation{\INAF }

\author[0000-0001-9472-041X]{Alec J.~M. Thomson} 
\affiliation{\CSIROBentley }

\author[0000-0003-0932-3140]{Andrea Bracco}
\affiliation{\INAFOAA }
\affiliation{\LPENS }

\author{D.~A. Del Rizzo}
\affiliation{\DRAO}

\author[0000-0003-2469-1611]{Ryan R. Ransom}
\affiliation{\OC }
\affiliation{\DRAO }

\author[0000-0003-2391-8650]{Robert I. Reid}
\affiliation{\Mayo}



\begin{abstract}
Structures in the magnetoionic medium exist across a wide range of angular sizes owing to large-scale magnetic fields coherent over the Galactic spiral arms combined with small-scale fluctuations in the magnetic field and electron density resulting from energy injection processes such as supernovae. For the first time, we produce diffuse Galactic synchrotron emission Faraday rotation maps covering all spatial scales down to $3'$ resolution for magnetic field studies. These maps complement total and polarized intensity maps combining single-antenna and interferometric data that have been produced, such as the Canadian Galactic Plane Survey (CGPS). Combined maps have sensitivity to large scales from the single-antenna component and angular resolution from the interferometric component. We combine Global Magneto-Ionic Medium Survey High-Band North single-antenna and CGPS aperture-synthesis polarization data after spatial filtering, producing Stokes $Q$ and $U$ maps for the four CGPS frequency channels. We calculate rotation measures (RMs) for all pixels using a linear fit to polarization angle versus wavelength squared. Smooth polarized emission regions require the large-scale sensitivity of the single-antenna to illuminate the Faraday rotation, while aperture synthesis reveals small-scale RM variability. While these maps show magnetic field structures on the full range of spatial scales they probe, the RM values should be interpreted with caution, as the narrow $\lambda^2$ coverage limits sensitivity to Faraday complexity. Despite this limitation of the CGPS 35 MHz bandwidth, we demonstrate that useful Faraday rotation information can be obtained from the combined dataset, highlighting the important synergy between future broadband interferometric and single-antenna polarization surveys.
\end{abstract}



\section{Introduction}\label{sec:intro}
The most commonly used method for studying the line-of-sight (LOS) component of the magnetic field in the interstellar medium (ISM) at radio frequencies relies on Faraday rotation. Faraday rotation changes the angle of the plane of polarization of linearly polarized radiation as
\begin{equation}
	\frac{\Delta\tau}{[\text{rad}]}=0.81\frac{\lambda^2}{[\text{m}^2]}\int_{\text{src}}^{\text{obs}} \frac{n_{\text{e}}}{[\text{cm}^{-3}]}\frac{B_{||}}{[\mu G]}\frac{dl}{[\text{pc}]}=\frac{\lambda^2}{[\text{m}^2]}\frac{\text{RM}}{[\text{rad m}^{-2}]},
    \label{eq:RM}
\end{equation}
where $n_e$ is the electron density along the LOS, $B_{||}$ is the LOS component of the magnetic field, and $dl$ is the LOS path-length increment \citep[see, e.g.][for details]{Ferriere_2021}. For a discrete radiating source, from which the polarized emission undergoes Faraday rotation as it propagates from the source (src) toward the observer (obs) through the intervening magnetoionized medium, there is a linear relationship between the change in polarization angle ($\Delta\tau$) and wavelength ($\lambda$) squared. The slope is then the single parameter known as the rotation measure (RM) describing that LOS. The relative simplicity of this specific scenario has been put to extensive use in attempts at mapping out the magnetic structure of the ISM \citep[e.g.][]{Simard-Normandin_1979, Rand_1994, Brown_2001, Brown_2007, VanEck_2011, Tahani_2018} as well as magnetic fields of external galaxies and the far more weakly magnetized intergalactic medium \citep[e.g.][]{Carretti_2022,Carretti_2023}. In these surveys, distant, compact sources, typically pulsars or active galactic nuclei (AGNs) of external galaxies, which are unresolved in the beam of the telescope and can therefore be treated as point sources, have been used as probes of the intervening medium.

In the past decade, with technological developments allowing for broadband radio polarization surveys with closely spaced frequency channels, there has been increased interest in mapping Galactic magnetism and polarization properties of the ISM through observations of the diffuse synchrotron emission from within the Galactic volume, which extends along each LOS, mixing with the Faraday rotation. In these scenarios, a single RM is not always representative of the LOS, and Equation~\ref{eq:RM} is generalized to allow for multiple RMs (called `Faraday depths') originating at different `source' locations along the LOS. In these cases a Faraday depth spectrum, derived from applying `RM synthesis' to broadband observations \citep{Burn_1966,Brentjens_2005} can provide information on the complexity and extent of the resulting RM structures.

The Global Magneto-Ionic Medium Survey (GMIMS), loosely comprising six component surveys, has been paving the way for full-sky, broadband, single-antenna polarization maps to which the technique of RM synthesis can be applied. To date, three GMIMS surveys have been published. In the Southern Hemisphere, Murriyang, CSIRO's Parkes 64 m radio telescope, has produced GMIMS-Low-Band South (GMIMS-LBS), covering 300-480~MHz \citep{Wolleben_2019}, and the Southern Twenty Centimeter All-sky Polarization Survey (STAPS), which contributes GMIMS-High-Band South (GMIMS-HBS), covering 1300-1800~MHz \citep{Sun_2025}. In the Northern hemisphere, the Dominion Radio Astrophysical Observatory (DRAO) John A. Galt telescope has produced the 1300-1800~MHz GMIMS-High-Band-North (GMIMS-HBN) survey \citep{Wolleben_2021}. Observations are underway for the POSSUM EMU GMIMS All Stokes UWL Survey (PEGASUS; Carretti et al., in preparation), covering 704-1440~MHz and contributing GMIMS-Mid-Band South. In the Northern Hemisphere, data from the Canadian Hydrogen Intensity Mapping Experiment \citep[CHIME;][]{chime_overview} and the DRAO GMIMS of the Northern Sky (DRAGONS) survey with the DRAO 15-m telescope (Ordog et al. in preparation) will contribute interferometric (aperture synthesis) and single-antenna components respectively to GMIMS-Low-Band North in the 350-1030~MHz range. The CHIME polarization data alone have already been put to use in the study of a 10$\arcdeg$~-~wide Faraday rotation feature \citep{Mohammed_2024}, and the DRAGONS polarization cube is near completion. Independently, the LOFAR Two-meter Sky Survey (LoTSS-DR2) has contributed a low-frequency, interferometric approach, mapping out a significant portion of the northern sky, with RM synthesis maps presented in \cite{Erceg_2022,Erceg_2024}.

A number of studies have begun to compare Faraday depths\footnote{Here we use `Faraday depth' to generalize either RMs or values derived from RM synthesis and Faraday depth spectra.} of compact sources and diffuse synchrotron emission in an effort to understand the differences between the distances probed by these data sets, and the distributions of thermal electrons and synchrotron-emitting regions that could give rise to the observed ratios of compact source to diffuse emission Faraday depth. \cite{Ordog_2017} and \cite{Ordog_2019} compared extragalactic source RMs to the diffuse emission RMs in their vicinity using aperture-synthesis data from the Canadian Galactic Plane Survey (CGPS). They concluded that it was possible to trace structures such as the large-scale magnetic field reversal in aperture-synthesis data alone, and that the aperture-synthesis RMs generally agree well with the compact source RMs in terms of the patterns traced as a function of Galactic longitude. \cite{Dickey_2019} compared first-moment (M1) maps of the GMIMS-HBN and GMIMS-LBS Faraday depth spectra to the \cite{Oppermann_2015} extragalactic RM map, as well as pulsar RMs, and found that GMIMS-HBN samples to larger distances along the LOS than GMIMS-LBS. In \cite{Dickey_2022} a comparison between GMIMS-HBN and the \cite{Hutschenreuter_2022} extragalactic RM map revealed an agreement in the Galactic longitude patterns traced by the two datasets at intermediate Galactic latitudes. \cite{Erceg_2022} also compared the LoTSS-DR2 M1 maps to the \cite{Hutschenreuter_2022} grid and found a correlation between the maps, while highlighting the level of Faraday complexity present in the low-frequency diffuse emission of LoTSS-DR2 that shows the deviations from a Burn slab \citep{Burn_1966} not discernible at higher frequencies. In all of these studies, the diffuse emission Faraday depths were derived from either single-antenna observations alone \citep{Dickey_2019,Dickey_2022} or interferometric observations alone \citep{Ordog_2017,Ordog_2019,Erceg_2022,Erceg_2024}, which may not provide the complete picture in all cases, and in some scenarios it may produce incorrect results \citep[e.g.][Appendix]{Gaensler_2001}.

In observations of total intensity, for both continuum and spectral line emission, it has been standard practice to combine aperture-synthesis and single-antenna data \citep[for a review and comparison of methods, see][]{Plunkett_2023}. In general, this produces maps with sensitivity to all spatial scales, from the largest scales present on the sky down to the limit set by the angular resolution of the aperture-synthesis telescope. The first and to-date only large survey to accomplish this addition with polarization data is the CGPS \citep{Landecker_2010}. We refer to such a dataset as a ``full-scale'' polarization dataset. The data published by \citet[hereafter L10]{Landecker_2010} brought together aperture-synthesis data with four frequency channels with single-antenna data covering only one channel because that was the only single-antenna dataset available at the time. Across multiple frequencies, channel-by-channel inclusion of the single-antenna component is also important for correctly determining spectral indices. In the case of Stokes $Q$ and $U$ data cubes for use in Faraday rotation analyses, the impact can be even more significant, potentially resulting in incorrect polarization angles if single-antenna data are not included, which in turn can lead to incorrect RM values. 

Combining aperture-synthesis and single-antenna data for the purpose of full-scale Faraday rotation studies of the diffuse polarized Galactic emission requires the two datasets to overlap in sky coverage, frequency coverage, and spatial scales. In the Northern Hemisphere a broadband application of this technique will be possible in the near future by combining CHIME data with GMIMS-LBN (DRAGONS) single-antenna data, though the resolution of this will be limited by CHIME's 100 m maximum baseline to $\sim 20'$. At the DRAO, an upgrade to the Synthesis Telescope (ST) has begun, which will ultimately provide broadband, arcminute-resolution interferometric data to complement all Northern-hemisphere GMIMS components and future single-antenna surveys. For the Southern Hemisphere, full-scale polarization datasets will be possible by combining the interferometric data from the Polarization Sky Survey of the Universe's Magnetism \citep[POSSUM;][]{Gaensler_2010} with the PEGASUS single-antenna survey (Carretti et al., in preparation).

Here we present a precursor to the type of full-scale polarization dataset that will be available with the completion of GMIMS, POSSUM, and the upgraded DRAO ST. We combine the four-channel diffuse polarization data from the CGPS that were used in \cite{Ordog_2017,Ordog_2019} with the data in an equivalent set of frequency channels from GMIMS-HBN. We calculate RMs for each pixel from linear fits to polarization angle versus wavelength squared on this combined, full-scale dataset. We use RMs as the best estimate of the Faraday depth for each LOS rather than Faraday depth spectra derived from RM Synthesis \citep{Brentjens_2005} because we are limited to four frequency channels. At 1420~MHz we expect little Faraday complexity, making linear fit RMs a reasonable approach. Even in the case of a Burn slab \citep[uniformly mixed synchrotron emission and Faraday rotation;][]{Burn_1966} the linear fit RM will match the mean value of Faraday depths from the slab. We highlight the usefulness of the technique and the structures that can be seen with only four frequency channels used to constrain the RMs, equivalent to a broad Rotation Measure Spread Function (RMSF) in the RM synthesis formalism. We compare the resulting map to the CGPS-only diffuse emission RMs, the CGPS point-sources RMs, and the Faraday depth maps derived from applying RM Synthesis to the full frequency coverage of the GMIMS-HBN data. The important result of this paper is the first full-scale polarization dataset in which we can begin to explore the Faraday rotation effects of the diffuse Galactic emission at arcminute resolution.

We present single-antenna and aperture-synthesis data used in Section~\ref{sec:data} and the methods we used to combine them in Section~\ref{sec:steps}. We show the resulting maps of polarized emission and RM in Section~\ref{sec:results} and discuss our findings and recommendations for future surveys in Section~\ref{sec:discussion}.
\newpage
\section{The Data}\label{sec:data}
Here we describe the datasets used in this work: the aperture-synthesis data from the CGPS and some additional DRAO ST fields, and the GMIMS-HBN cube. 

\subsection{DRAO Synthesis Telescope data}\label{sec:data_cgps}
The aim of the CGPS was to map the key ISM constituents with $1'$ angular resolution in the portion of the Galactic plane visible from the Northern Hemisphere. This included polarized and total continuum emission at 1420 MHz, 21-cm H{\textsc{i}} spectral line emission, and total intensity continuum emission at 408~MHz. The survey was completed between 1995 and 2010, and the data were observed in three phases covering different longitude ranges: (i) $74\arcdeg<\ell<147\arcdeg$ \citep{Taylor_2003}, with a focus on the survey setup and a description of the data processing routines; (ii) expanded to $66\arcdeg<\ell<175\arcdeg$ with a high-latitude extension up to $b = 17.5\degr$ in the longitude range $101\arcdeg<\ell<116\arcdeg$ \citep{Landecker_2010}, concentrating on the linear polarization component; and (iii) further expanded to $52\arcdeg<\ell<193\arcdeg$ \citep{2017AJ....154..156T}. In this final paper the 408~MHz survey was released. At 1420~MHz the survey (not including two latitude extensions) covers $-3\arcdeg<b<5\arcdeg$ in latitude along the full longitude range. At 408~MHz the latitude coverage is $-6.5\arcdeg<b<8.5\arcdeg$.

The high-resolution, interferometric component of the survey (which is the dataset we use in this work) consists of observations with the DRAO ST. For the 1420~MHz continuum emission, the ST observed in four 7.5 MHz bands within a 35~MHz window, centered at the 1420~MHz H{\textsc{i}} line \citep{Landecker_2010} in order to avoid radial smearing of the synthesized beam arising from imprecise delay compensation \citep{Elsmore_1966}. The angular resolution is 58$^{\prime\prime}\times58^{ \prime\prime}$cosec($\delta$) using natural weighting of the baselines, where $\delta$ is the declination. At 1420~MHz, the field of view is 107$^{\prime}$ in diameter at 50\% power. The survey specifications are summarized in Table~1 in L10.

Following initial processing, the four frequency channels were combined for each field, and based on the primary beam attenuation pattern \citep{Taylor_2003} a weighting map was used to combine the ST fields into $5.12\arcdeg\times5.12\arcdeg$ mosaics. These published 1420~MHz CGPS Stokes $I$, $Q$ and $U$ data products are available from the Canadian Astronomy Data Centre (CADC). 

A significant contribution from the CGPS survey was a compact-source RM catalog that was produced by using the four ST bands \textit{separately} in order to allow for linear fits to polarization angle as a function of $\lambda^2$ on emission from unresolved sources (primarily AGNs) within the dataset \citep{Brown_2003b,VanEck_2021}. \citet{Ordog_2017,Ordog_2019} later demonstrated that RMs for the diffuse emission, calculated from the four bands, revealed the large-scale Galactic magnetic field just as effectively as RMs derived from compact sources.

In this work, we also use the four separate channels of CGPS ST 1420~MHz data, beginning with the individual archived fields (separate pointings, without primary beam correction) rather than the published mosaics. The main portions of the CGPS, covering the Galactic disk from $\ell=52\arcdeg$ to $\ell=193\arcdeg$, consist of 384 of these fields. We also include the 14 `Ordog-Brown' fields, adjacent to the CGPS, that were observed as a follow-up to the diagonal field reversal analysis in \cite{Ordog_2017}, as well as 5 additional fields from the ST archive. These 19 fields cover approximately $64\arcdeg\leq\ell\leq74\arcdeg$, $5\arcdeg\leq b \leq8\arcdeg$ and $52\arcdeg\leq\ell\leq62\arcdeg$, $-6\arcdeg\leq b \leq -3\arcdeg$.

\subsection{GMIMS-HBN}\label{sec:data_gmims}
The component of GMIMS that overlaps both spatially and in frequency with the CGPS-ST 1420~MHz data is GMIMS-HBN, observed with the DRAO John A. Galt 26 m telescope. Left- and right-hand circularly polarized inputs (L and R) allow for reliably determined Stokes $Q$ and $U$ products from the cross-correlations, LR* and RL*, respectively. Details about the telescope, receiver, observing method, and calibration are provided in \cite{Wolleben_2010,Wolleben_2021}. The published data cubes consist of Stokes $I$, $Q$ and $U$ in 409 frequency channels spaced by 1.184~MHz, spanning 1280 to 1750 MHz. GMIMS-HBN covers a declination range of $-20\arcdeg$ to $89\arcdeg$ with a spatial resolution of $40'$. The 26 m diameter of the reflector means the data are reliable out to $\sim$9~m, corresponding to the half-power level in terms of $uv$-plane baseline coverage \citep{Landecker_2010}. As outlined in \cite{Wolleben_2021}, a Faraday depth cube, derived from applying RM Synthesis using the CIRADA RM-tools package \citep{Purcell_2020}, has also been published, and we make use of this cube for comparison with the aperture-synthesis-only and the combined full-scale RM values.

GMIMS-HBN has been corrected empirically for instrumental polarization (conversion of Stokes $I$ into $Q$ and $U$) by iteratively adjusting Stokes $I$ leakage coefficients in the Galactic center region to minimize the spurious polarization, which was initially estimated to be $3\%$ of the total intensity \citep{Wolleben_2021}. These frequency-dependent leakage terms were then applied to the entire map and led to a factor of 10 reduction in instrumental polarization in regions of bright total intensity. In the published GMIMS-HBN maps, there is a small amount of Stokes $I$ to $Q$ and $U$ leakage remaining in the Galactic disk. However, this is strongest in the inner Galaxy, particularly below $\sim 60\arcdeg$ Galactic longitude, while the CGPS spans a higher longitude range, mostly confined to the outer Galaxy. Therefore, the effects of leakage on combining these datasets is likely minimal apart from the Cygnus X region, which we discuss in reference to the maps in Section~\ref{sec:results}. We do not attempt further refinement of instrumental polarization effects on the extended, diffuse regions in the map, as regions with low Stokes $I$ emission are subject to a high degree of uncertainty in their fractional polarization owing to the missing Stokes $I$ zero level in GMIMS-HBN. The small, $0.3\%$ remaining leakage reported in \cite{Wolleben_2021} allows for reliable polarization maps.

Other minor discrepancies exist in the GMIMS-HBN survey, and adjustments to the data processing may provide improvements in the future following a thorough investigation of the issues. Nevertheless, the improved spatial sampling and frequency coverage of GMIMS-HBN, compared to the \citet[hereafter W06]{Wolleben_2006} polarization survey used in the CGPS, make it the best dataset available in this frequency range, and it is well-suited to the experiments presented in this paper.

\section{Steps for combining datasets}\label{sec:steps}
The process for combining the datasets is based on the `feathering' method used in L10. In that work, full-band polarization data at 1420 MHz from the Effelsberg Medium-Latitude Survey \citep[EMLS;][Reich et al. in prep.]{Uyaniker_1998,Uyaniker_1999,Reich_2004}, and the DRAO John A. Galt single-antenna survey (W06), were combined with the aperture-synthesis data from the DRAO ST. In the analysis presented here, we cannot use the EMLS data because the separate frequency channels required for RM calculations do not exist. Here we outline the steps in the procedure and provide examples of the resulting maps. We used the DRAO \texttt{export package} software for all processing steps \citep{Higgs_1997}.

\begin{table}
	\centering
	\begin{tabular}{|c|c|c|}
		\hline
		\hline
		\textbf{DRAO ST} & \textbf{DRAO ST}  & \textbf{GMIMS-HBN} \\
		\textbf{Band} & \textbf{$\nu$ range (MHz)} & \textbf{$\nu$ channels (MHz)} \\ \hline
		A &1403.14 - 1411.01 &1404.58 - 1409.32 \\ \hline
		B &1410.08 - 1417.95 &1410.50 - 1416.42 \\ \hline
		C &1423.66 - 1431.54 &1424.71 - 1430.63 \\ \hline
		D &1430.57 - 1438.44 &1431.81 - 1437.74 \\ \hline
		\hline
	\end{tabular}
	\caption{DRAO ST frequency ($\nu$) ranges for each channel (A-D) and the corresponding selected GMIMS-HBN frequency channels used to create DRAO ST-equivalent bands. Note that the edges of the bands do not match exactly owing to the differing channelization of the two data sets but this has a negligible impact on the results.}
	\label{gmims4bands}
\end{table}

\begin{figure*}
	\centering
	\includegraphics[width=0.8\textwidth]{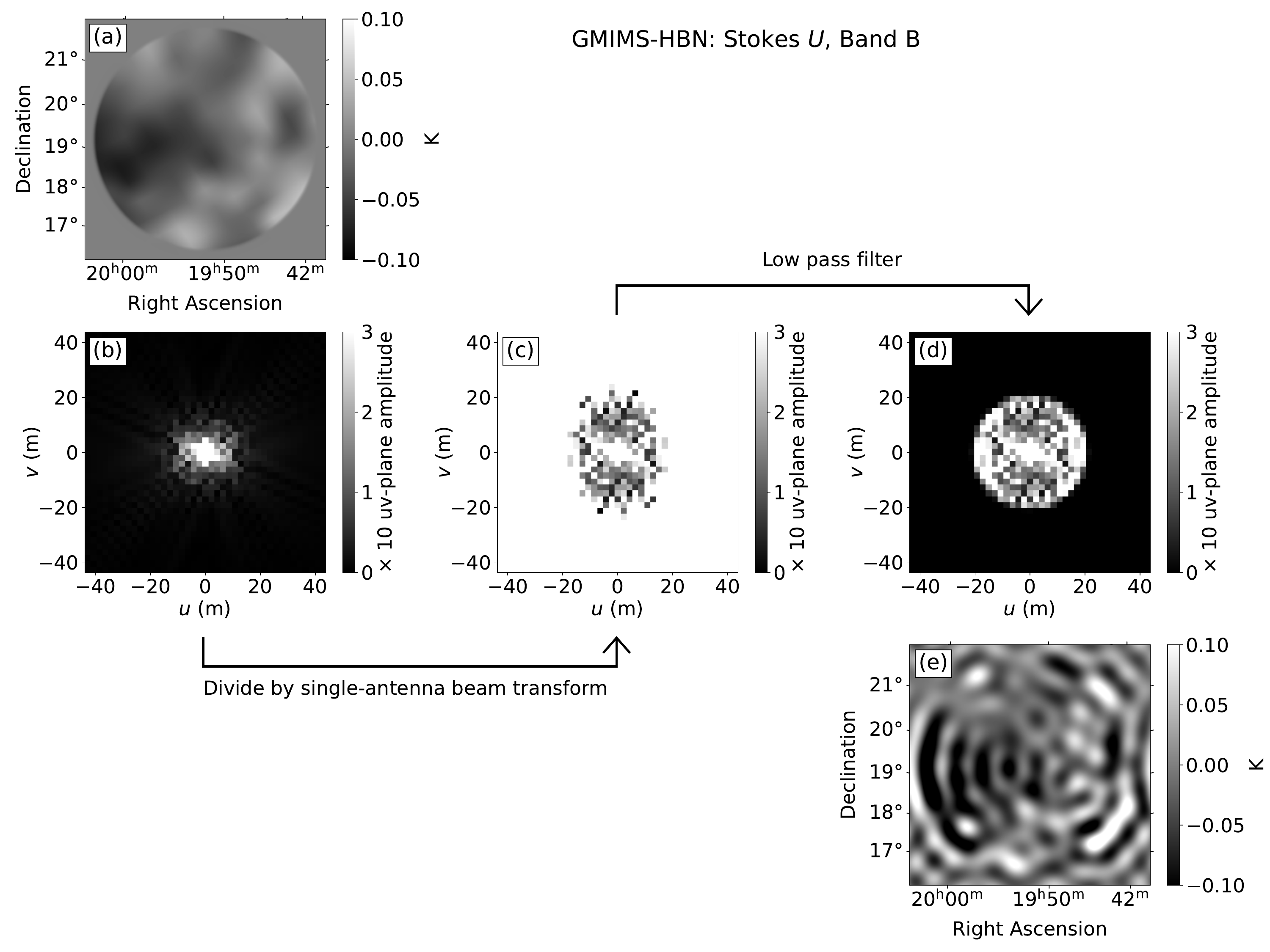} 
	\caption{The steps in preparing a GMIMS-HBN map to be combined with the corresponding DRAO ST field. This example shows Stokes $U$ for band B of one of the Ordog-Brown fields (OB12). (a) Tapered GMIMS-HBN map, regridded to match DRAO ST field grid spacing; (b) the GMIMS-HBN field transformed to the $uv$-plane; (c) the visibilities divided by the single-antenna beam transform to deconvolve the beam from the image; (d) the deconvolved visibilities low-pass-filtered to remove the unusable high spatial frequency components; (e) the resulting visibilities transformed back to the image plane.
	}
	\label{combiningfirststeps}
\end{figure*}

\subsection{Preparing the GMIMS-HBN Single Antenna data}\label{sec:steps_SA}
The procedure begins with tailoring the GMIMS-HBN single-antenna dataset to complement the DRAO ST aperture-synthesis dataset. The following steps are necessary owing to the differences in: the observing methods of the two datasets, the gridding of the maps in each, and their frequency coverage and spacing.

\begin{enumerate}[label=(\roman*)]
\setlength\itemsep{1em}
\item \textit{Make GMIMS-HBN frequency bands}: We generated four Stokes $Q$ and $U$ maps of GMIMS-HBN corresponding to the DRAO ST frequency bands by averaging the GMIMS-HBN channels that lie within $\pm$3.5 MHz of the center of each CGPS band. The GMIMS-HBN channels included in each band are shown in Table~\ref{gmims4bands}. The 7.5~MHz bandwidth of the DRAO ST bands places the edges of the bands slightly outside the corresponding GMIMS-HBN band edges, and offsets between the band edges vary slightly from band to band owing to the differing channelization of the two datasets. This has a negligible impact on the results, as the difference in bandwidth depolarization is small compared to other uncertainties in this experiment. 
\item \textit{Regrid and taper GMIMS-HBN fields}: We regridded the four-band GMIMS-HBN Stokes $Q$ and $U$ maps to match the ST image grid spacing of 0.333$^{\prime}$ in equatorial coordinates. We then divided these into regions corresponding to the centers of the CGPS and Ordog-Brown ST fields, with corresponding sizes (512 $\times$ 512 pixels for CGPS; 1024 $\times$ 1024 pixels for Ordog-Brown), and tapered them to approximate the ST field of view (primary beam). An example for Stokes $U$, band B, is shown in Figure~\ref{combiningfirststeps}a.
\item \textit{Deconvolve single-antenna beam in $uv$-plane and low-pass filter}: The GMIMS-HBN data need to be corrected for the beam before they can be added to the DRAO ST data. The observed image is the true image convolved with the primary beam of the telescope:
\begin{equation}
	I_{\text{obs.}} = I_{\text{true}} \ast \text{beam}.
\end{equation}
In order to flatten the single-antenna beam response in the $uv$-plane (and thereby approximately recover $I_{\text{true}}$) for use in the feathering step, we must deconvolve the GMIMS-HBN beam from the image. This can be done by performing the Fourier transform to the $uv$-plane, and dividing the resulting visibilities by the Fourier transform of the beam \citep{Landecker_2010}:
\begin{equation}
\mathcal{F}(I_{\text{true}})=\frac{\mathcal{F}(I_{\text{obs.}})}{\mathcal{F}(\text{beam})}
\end{equation}
To model the GMIMS-HBN Fourier-transformed beam, we used a Gaussian with a half-width at half-maximum of 9~m in the $uv$-plane, which corresponds to the 40$^{\prime}$ spatial resolution of the 26 m single-antenna telescope. We then low-pass-filtered the modified GMIMS-HBN visibilities in the $uv$-plane with a maximum of 18~m, slightly beyond the final extent of the data that is ultimately used. The deconvolution and low-pass filtering are shown in Figure~\ref{combiningfirststeps}b-d.
\item \textit{Taper to match primary beam of ST}: After low-pass filtering, we transformed the GMIMS-HBN visibilities back to the image plane (Figure~\ref{combiningfirststeps}e), and tapered the resulting image to match the frequency-dependent primary beam of the ST, which is fitted by a cos$^6$ function with a FWHM of 107.2$^{\prime}$. The tapered GMIMS-HBN image is shown in Figure~\ref{combiningsecondsteps}a.
\end{enumerate}

\subsection{Combining the datasets}\label{sec:steps_combine}
In order to directly combine the single-antenna and aperture-synthesis datasets, both must be weighted appropriately in the overlap region of the $uv$-plane. We transformed both the DRAO ST data and the GMIMS-HBN data (beam-matched to the ST) back to the $uv$-plane to perform this weighting, known as feathering. We weighted the DRAO ST data from 0 at 8.572~m, which is slightly below the nominal shortest useful baseline for the ST (12.858~m), to 1 at 17.144~m, and weighted the GMIMS-HBN data from 1 at 8.572~m to 0 at 17.144~m, using a cubic weighting function. The boundaries were selected as integer multiples of the spacing between adjacent elliptical tracks in the $uv$-plane for the DRAO ST. This feathering of the DRAO ST and GMIMS-HBN yields an overlap region of width 8.572~m, which corresponds to twice the DRAO ST $uv$-plane spacing. Given the limitations of the $uv$-plane coverage of the two datasets, this is the best that can be done for the combination. 
\begin{figure}
	\centering
	\includegraphics[width=\hsize]{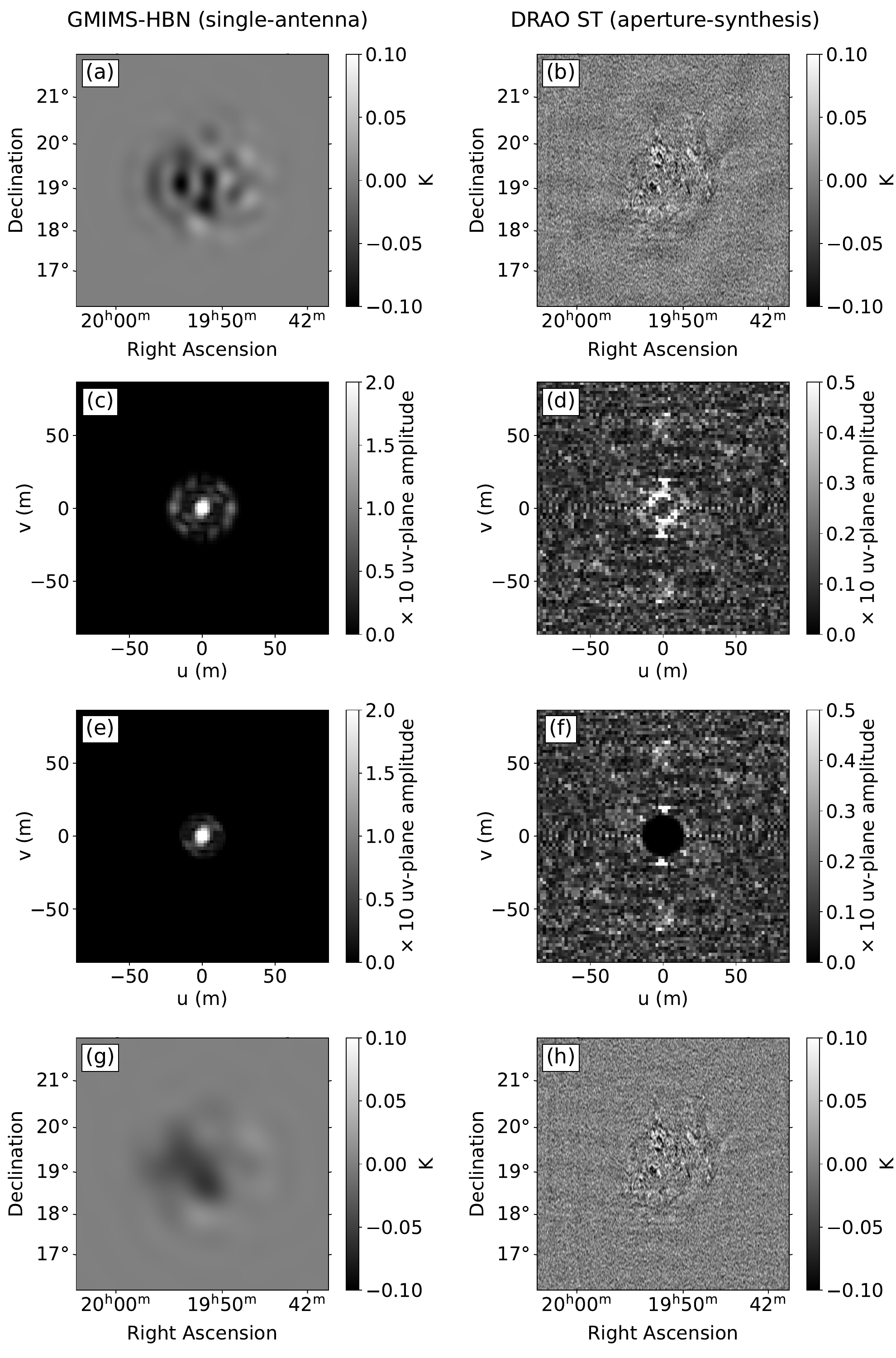} 
	\caption{The steps in feathering GMIMS-HBN and DRAO ST visibilities in order to combine the datasets. This example shows Stokes $U$ for band B of field OB12. (a) The deconvolved, low-pass-filtered GMIMS-HBN map tapered to match the primary beam of the DRAO ST; (b) the original DRAO ST field; (c) GMIMS-HBN transformed to the $uv$-plane; (d) DRAO ST transformed to the $uv$-plane; (e) the GMIMS-HBN visibilities feathered to discard high spatial frequencies; (f) the DRAO ST visibilities feathered to discard low spatial frequencies; (g,h) the GMIMS-HBN and DRAO ST respectively, transformed back to the image plane, ready to be added together.
	}
	\label{combiningsecondsteps}
\end{figure}

After transforming the feathered visibilities of both datasets back to the image plane, we added them together to form the final combined image. This procedure, which is illustrated with the examples in Figure~\ref{combiningsecondsteps}, was repeated for all  CGPS and Ordog-Brown fields, which could then be combined into mosaics as described in the following section.

\begin{figure*}
	\centering
	\includegraphics[width=0.8\textwidth]{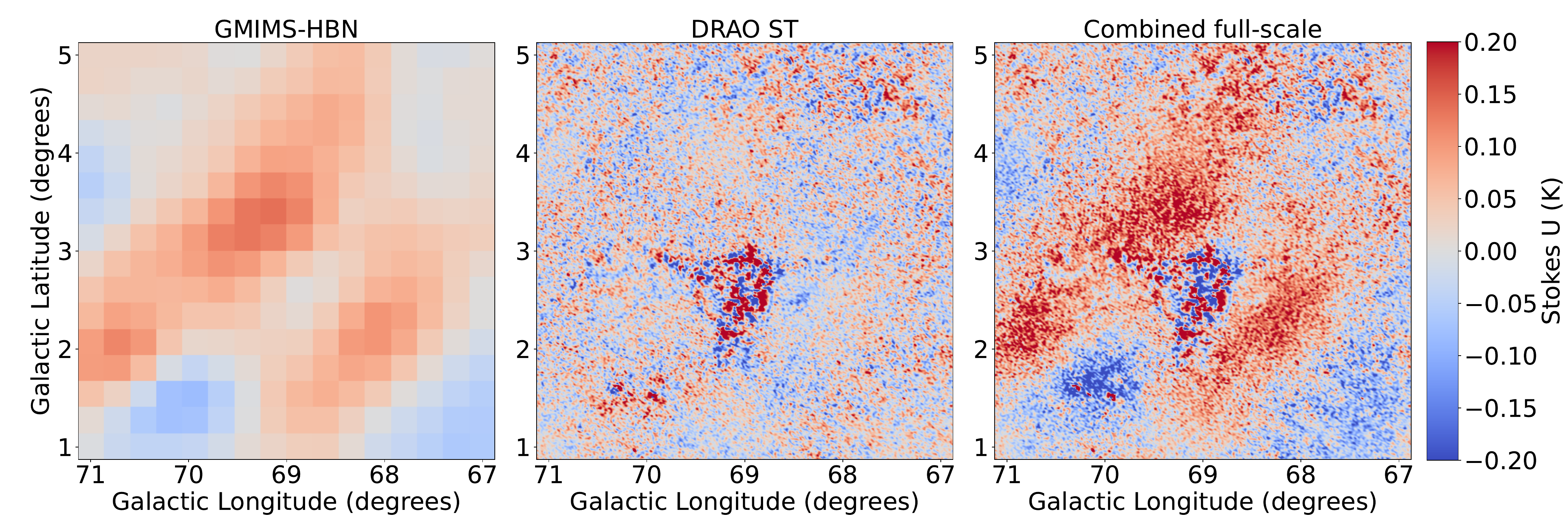} 
	\caption{Sample mosaics of GMIMS-HBN (left), DRAO ST (middle) and combined data (right) for Stokes $U$, band B. Regions of bright positive emission covering a few square degrees that are filtered out by the ST are incorporated from the GMIMS-HBN data.
	}
	\label{combinedmosaicexample}
\end{figure*}

\subsection{Mosaics and full-sky maps}\label{sec:steps_mosaics}
After converting the fields to Galactic coordinates (and rotating the polarization angles accordingly), we combined sets of fields into $4.5\arcdeg \times 4.5\arcdeg$ mosaics using the \texttt{supertile} routine from the DRAO \texttt{export package} \citep{Higgs_1997}, which applies a primary-beam-dependent weighting pattern to average the signal between overlapping fields. We followed the CGPS gridding and mosaic-centering convention in constructing the mosaics. An example of a mosaic, comparing the data from GMIMS-HBN and the DRAO ST with the combined data, is shown in Figure~\ref{combinedmosaicexample}. We then combined the resulting 78 overlapping mosaics using the \texttt{reproject} package in \texttt{Python}, yielding combined full-scale Stokes $Q$ and $U$ maps at each of the four frequencies. We further convolved these maps with a 3 arcmin Gaussian beam and down-sampled to $0.01\arcdeg$ grid spacing, degrading the resolution slightly but significantly reducing noise in the maps. We mask a 1$\arcdeg$ wide strip at $\ell=180\arcdeg$, as a discontinuity in the GMIMS-HBN data at this longitude produces artifacts in the Fourier transform steps described in Section~\ref{sec:steps_combine}. We do not mask other artifacts arising from either dataset, but we make note of these in discussing the maps in Section~\ref{sec:results}. Henceforth we refer to the combined single-antenna and aperture-synthesis Stokes $Q$, $U$ and derived RM maps as the ``full-scale'' data.

\subsection{Assessment of gap in $uv$ coverage}\label{sec:discussion_gap}
For adding short spacings to aperture-synthesis data in L10, Effelsberg 100 m single-antenna data were used to span the slight gap in $uv$-plane coverage between the 26 m and the ST. For the present case, the intermediate coverage is lacking and the parameters we selected in order to stitch together the GMIMS-HBN and DRAO aperture-synthesis data approach the reliable limits of both datasets. Here we briefly investigate the implications of this missing information.

\subsubsection{Radial $uv$ profiles of visibilities}\label{sec:discussion_radialuv}
\begin{figure}[h] 
	\centering
	\includegraphics[width=\hsize]{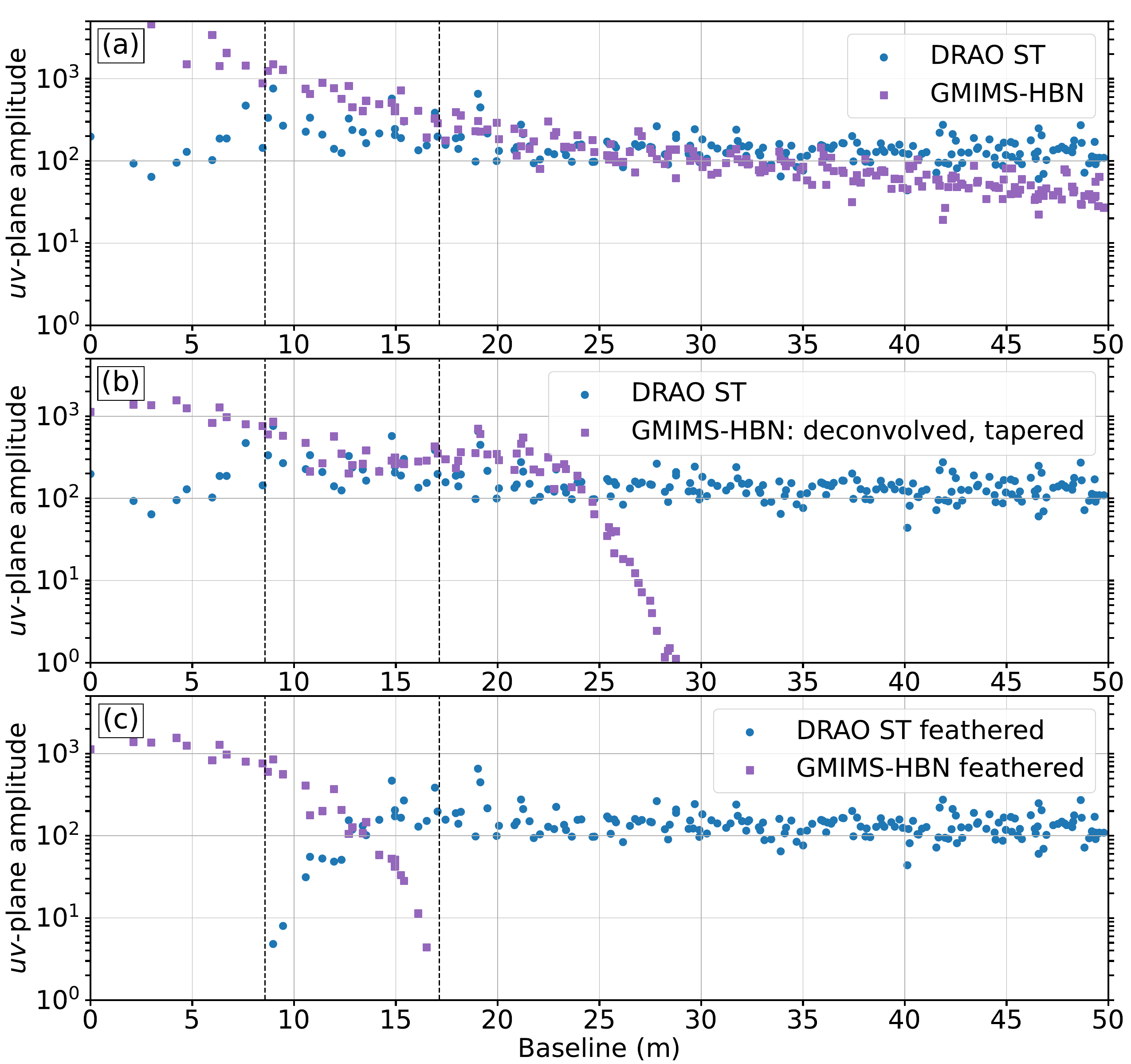} 
	\caption{Radial profile of GMIMS-HBN (purple squares) and DRAO ST data (blue circles) in the $uv$-plane. This example shows Stokes $U$ for band C of the Ordog-Brown field `OB12' (19h 52m, $19\arcdeg25'$). (a) The initial GMIMS-HBN and DRAO ST visibilities as a function of radial distance (baseline), $r=\sqrt{u^2+v^2}$, from the center of the $uv$-plane. Here the GMIMS-HBN data were matched to the ST gridding and tapered to approximate the ST field of view (Section~\ref{sec:steps_SA}, ii). (b) Visibilities of GMIMS-HBN data with the beam deconvolved and the image tapered to the ST primary beam (Section~\ref{sec:steps_SA}, iv). The DRAO ST visibilities remain unchanged. (c) GMIMS-HBN and DRAO ST feathered versions (Section~\ref{sec:steps_combine}). The dashed vertical lines indicate the feathering region boundaries.
	}
	\label{fieldradialprofile}
\end{figure}

 We begin by plotting the radial profiles of the visibilities in the $uv$-plane for both the GMIMS-HBN and the DRAO ST for a sample ST field, as shown in Figure~\ref{fieldradialprofile}. Panel (a) shows the GMIMS-HBN and DRAO ST visibilities, panel (b) shows the same but with the GMIMS-HBN matched to the DRAO ST beam, and panel (c) shows the feathered versions of both. The dashed vertical lines indicate the feathering region boundaries.

\begin{figure}[h] 
	\centering
	\includegraphics[width=\hsize]{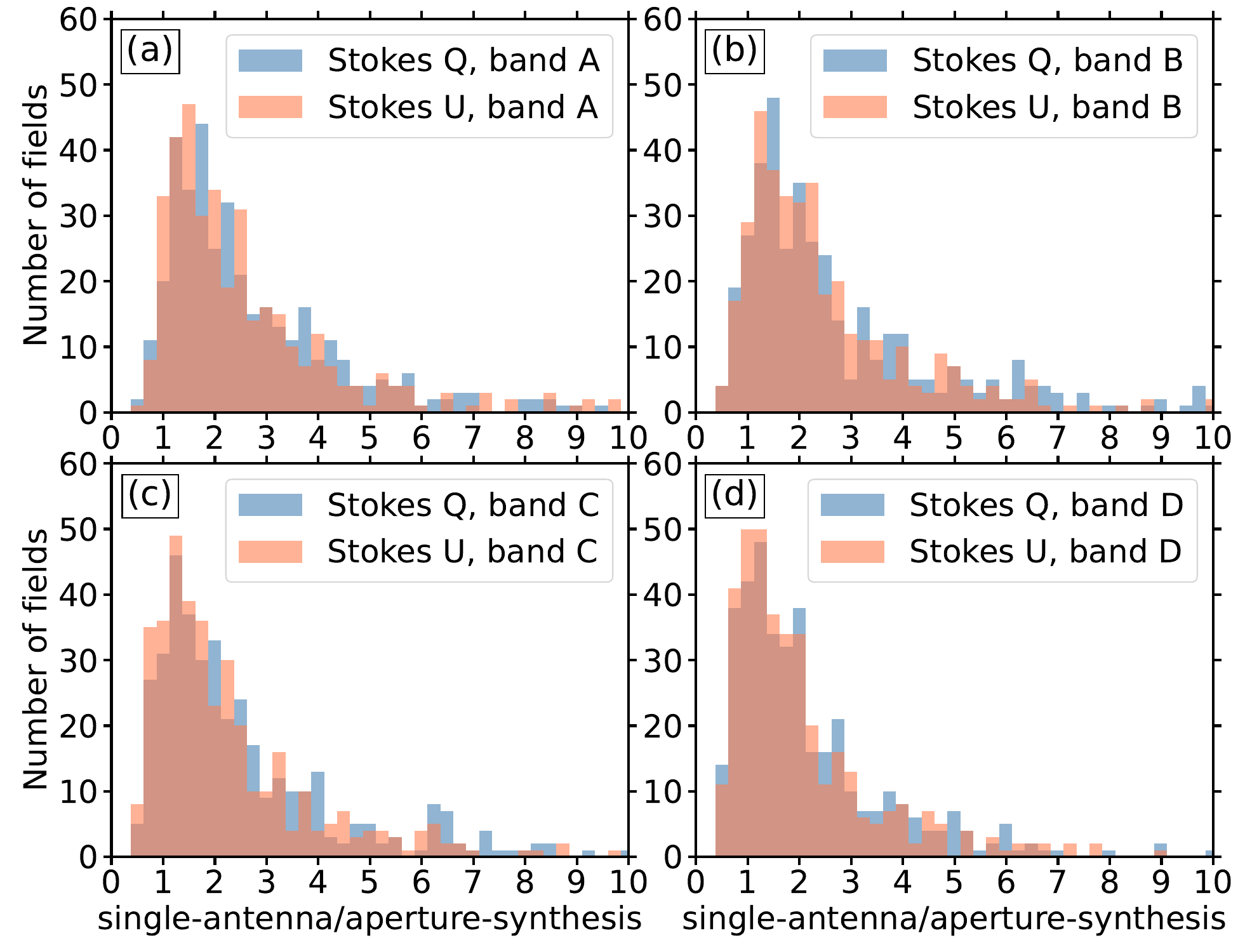} 
	\caption{The distribution of ratios of GMIMS-HBN to DRAO ST polarization data in the $uv$-plane. The ratio is calculated separately for each band of Stokes $Q$ and $U$ in the region of the $uv$-plane where the datasets of the 403 CGPS and Ordog-Brown fields used in this experiment are assumed to have overlapping coverage. The ratio should be close to unity for reliable signal at all baselines.
	}
	\label{uvplanehistogram}
\end{figure}

The deconvolved visibilities (Figure~\ref{fieldradialprofile}b) of the GMIMS-HBN data begin to rise slightly and to diverge from the DRAO ST visibilities beyond baselines of $\sim15$~m. This is because the Gaussian beam profile by which the visibilities are divided approaches zero with increasing distance from the center in the $uv$-plane. Equivalently, this is a consequence of using data from the single antenna significantly beyond its half-power sensitivity in the $uv$-plane. For the particular field shown in Figure~\ref{fieldradialprofile}, the range in radial distance over which GMIMS-HBN and the DRAO ST agree extends down to $\sim10$~m, below the minimum spacing of the ST antennas. This is because observing at the low declination ($19\arcdeg25'$ for the Ordog-Brown field `OB12') leads to foreshortening of the baselines and allows for lower spatial frequencies to be observed with the ST. The lack of foreshortening at higher declinations means that we may be introducing more noise by using these short baselines.

The distributions of ratios between the beam-corrected GMIMS-HBN and the DRAO ST in the overlap region for Stokes $Q$ and $U$ in the four separate bands are shown in Figure~\ref{uvplanehistogram}. Ideally, these ratios should be close to unity for meaningful contributions from both the single-antenna and aperture-synthesis data in the overlap baseline range. An average of 1.15$\pm0.2$ was reported in L10 for the case of combining CGPS DRAO ST data with Effelsberg data. In our case, although the distributions of the ratios peak between 1 and 1.5, the spread in the ratios is significantly higher, and the GMIMS-HBN values tend to be larger than the DRAO ST values, indicating the amplification of noise in the higher spatial frequencies of the single-antenna data. The resulting data products (mosaicked $Q$, $U$, PI and RM maps) may suffer from increased noise at the spatial frequencies corresponding to the problematic baselines. DRAO ST field pointings with particularly high ratios of single-antenna to aperture-synthesis data correspond to a few regions with very high PI; regions surrounding the bright sources Cassiopeia A, Cygnus A, and W3; and a discontinuity in the GMIMS-HBN data at $\ell=180\arcdeg$ producing artifacts in the Fourier transform steps described in Section~\ref{sec:steps_combine}. In our analysis in Section~\ref{sec:discussion_uses} we avoid these regions to prevent misinterpretations of the combined full-scale maps.

\subsubsection{Simulated observations}\label{sec:discussion_simulate}
\begin{figure*}
	\centering
	\includegraphics[width=0.8\textwidth]{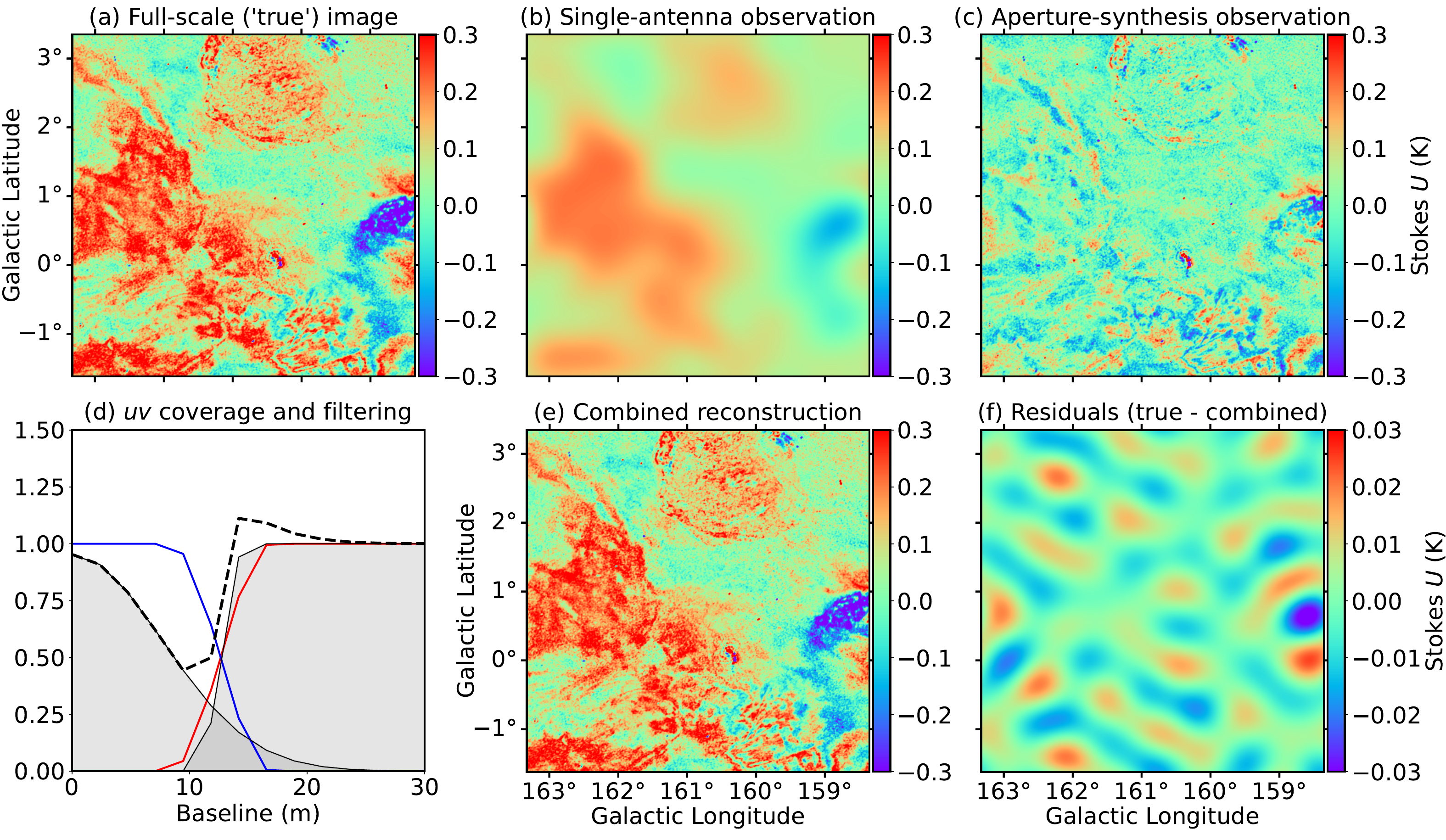} 
	\caption{Mock observations of a full-scale Stokes $U$ image with the GMIMS-HBN and DRAO ST telescope parameters, and the resulting combined image. (a) The full-scale image. (b) The simulated single-antenna image. (c) The simulated aperture-synthesis image. (d) The $uv$-plane coverage used in each of the single-antenna and aperture-synthesis mock observations (black lines with gray shading) along with the feathering functions by which we multiplied the simulated single-antenna (blue) and aperture-synthesis (red) data. (e) The resulting reconstructed image from combining the mock observations to produce the full-scale map. (f) The residuals for comparison with the input map.}
	\label{fig:simulation}
\end{figure*}

In order to simulate the effects of a gap in the $uv$-coverage between the single-antenna and aperture-synthesis data, we made mock observations using parameters corresponding to the GMIMS-HBN and the DRAO ST and then applied the procedure described in Sections~\ref{sec:steps_SA}~and~\ref{sec:steps_combine} to combine them. Our starting image for this experiment is a combined CGPS, DRAO 26 m and Effelsberg 100 m Stokes $U$ mosaic from L10, which does \textit{not} have the gap in $uv$-coverage. This region covers an area $5\arcdeg\times 5\arcdeg$, selected because it includes structures on a range of spatial scales but does not include bright sources that would produce leakage artifacts in the single-antenna or aperture-synthesis observations (Figure~\ref{fig:simulation}a).

We simulated the single-antenna data (Figure~\ref{fig:simulation}b) by filtering this image in the $uv$ domain using a Gaussian with half-power at 9~m (Figure~\ref{fig:simulation}d; left shaded region). We simulated the aperture-synthesis data (Figure~\ref{fig:simulation}c) by high-pass filtering starting at 15~m (Figure~\ref{fig:simulation}d; right shaded region), slightly greater than the shortest nominal spacing of the DRAO ST. The roll-off of the filter is an approximation, assuming zero power below 11~m, slightly less than the shortest nominal spacing of the DRAO ST. Although this may not be the exact profile for the DRAO ST sensitivity, a visual comparison between the simulated aperture-synthesis map and the map of the same region with DRAO ST-only data shows good agreement. The black dashed line in Figure~\ref{fig:simulation}d illustrates the sum of the simulated sensitivities of the single-antenna and aperture-synthesis components. The range of $uv$-plane values where this sum is below 1 highlights the baseline coverage over which we have reduced sensitivity in the combined dataset.

Prior to applying the deconvolution step described in Section~\ref{sec:steps_SA} (iii), we injected noise into the simulated single-antenna data in the $uv$ domain, approximately equal to the power in the single-antenna data at the maximum baseline of 17.1~m used in the feathering step. The purpose of this was to simulate the effect of enhancing noise in the data at high spatial frequencies by dividing by the Fourier transform of the beam. We then feathered the simulated single-antenna and aperture-synthesis data as described in Section~\ref{sec:steps_combine}. The blue and red lines in Figure~\ref{fig:simulation}d, which add up to a value of 1 across all baselines, illustrate the feathering profiles. Finally, we combined the feathered data, resulting in the image shown in Figure~\ref{fig:simulation}e. The residuals between the original and the simulated combined maps are shown in Figure~\ref{fig:simulation}f.

Although there is some structure in the residuals map, the magnitudes of the differences are generally small. In this test, the distribution of the residuals has a standard deviation of 0.008~K, while the distribution of the full-scale map itself has a standard deviation of 0.1~K. This illustrates that the majority of the information is recovered even in the case of using data from the single-antenna component well beyond its ideal coverage in the $uv$ domain. Taking the ratio of these distribution widths as the fractional error introduced by the simulated process gives approximately $6\%$ error in Stokes $U$. Assuming the same error in Stokes $Q$ also yields the same error in PI, and propagates to an error of $\delta QU/2\text{PI}$ in polarization angle, where $\delta QU$ is the 0.008~K error in Stokes $Q$ and $U$. Sufficient signal in the PI map results in minimal error introduced in polarization angle.

We have explored the shortcomings of the combined maps arising from the slight gap in $uv$-plane coverage between the the two component datasets. We conclude that these deficiencies do not invalidate the experiment, and the resulting maps are adequate for a prototype demonstration of diffuse emission RM analysis from full-scale Stokes $Q$ and $U$ images.

\section{Results}\label{sec:results}
We show the RM and PI maps constructed from the Stokes $Q$ and $U$ maps described in Sections~\ref{sec:steps_SA}~to~\ref{sec:steps_mosaics} in Figures~\ref{fig:fullmap1}-\ref{fig:fullmap7}. Figures~\ref{fig:fullmap2}-\ref{fig:fullmap6} cover the same longitude ranges as Figures~5-7 in L10. Figures~\ref{fig:fullmap1} and \ref{fig:fullmap7} include the longitude extensions covered by Phase 3 of the CGPS, and Figure~\ref{fig:fullmap1} also includes the `Ordog-Brown' fields observed for a follow-up to the magnetic field reversal region study in \cite{Ordog_2017}.

\subsection{Polarized intensity maps}\label{sec:results_PImaps}
\begin{figure*}
    \centering
    \includegraphics[width=1\textwidth]{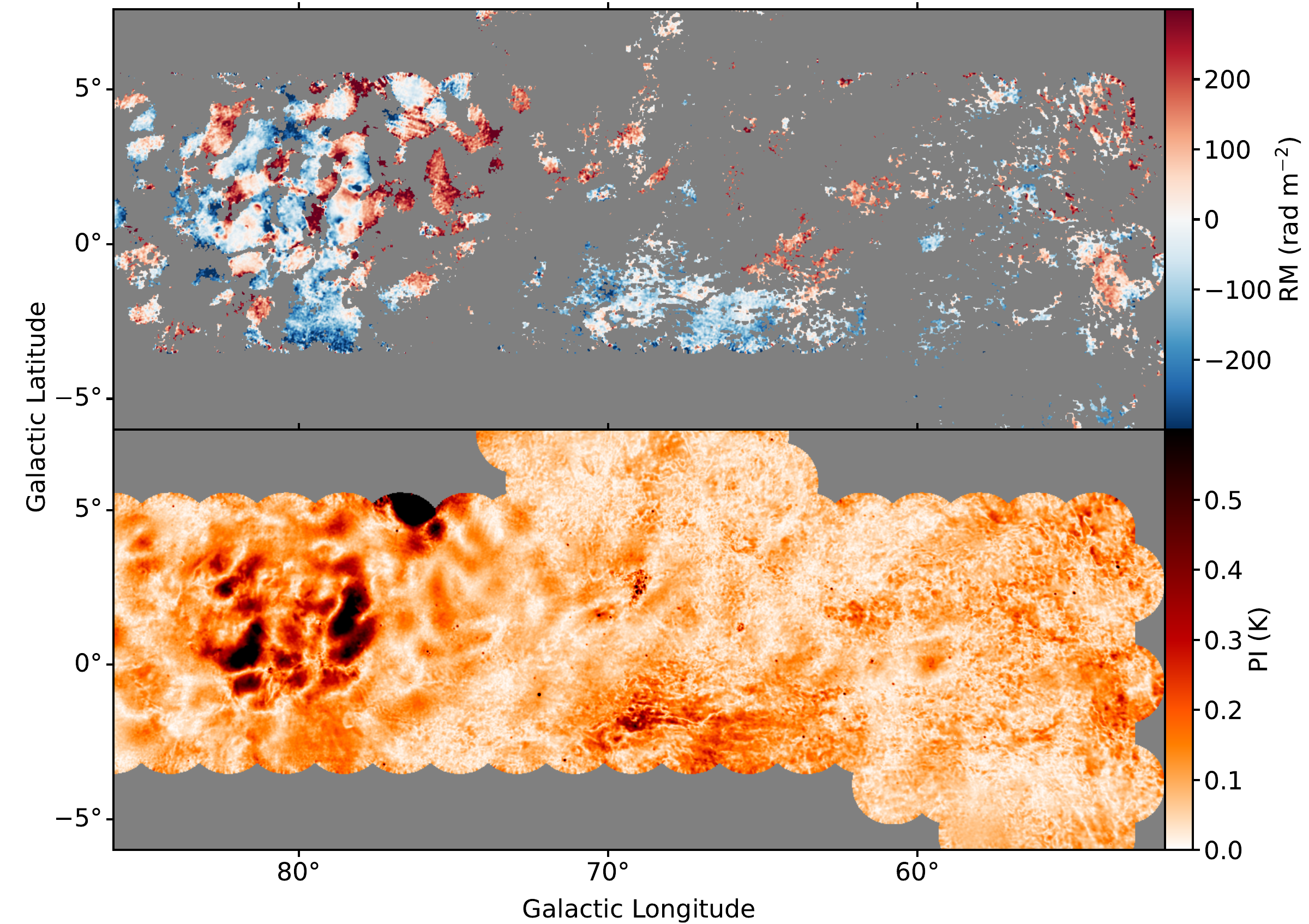}
    \caption{Combined full-scale RM and PI maps for $52\arcdeg<\ell<86\arcdeg$. Top: RM for pixels with PI$>0.1$~K (gray denotes the masked out regions). Bottom: PI calculated as $\text{PI} = \sqrt{\langle Q \rangle^2+\langle U \rangle^2}$ where the averages $\langle Q \rangle$ and $\langle U \rangle$ are taken over the four ST frequency channels. These maps include the region in the top panel of Figure~5 in L10, as well as previously unpublished CGPS Phase 3 data and the latitude extensions of the Ordog-Brown fields. Note the instrumental polarization in the Cygnus X region, $75\arcdeg<\ell<85\arcdeg$ arising from the GMIMS-HBN data.
    }
    \label{fig:fullmap1}
\end{figure*}

\begin{figure*}
	\centering
	\includegraphics[width=0.78\textwidth]{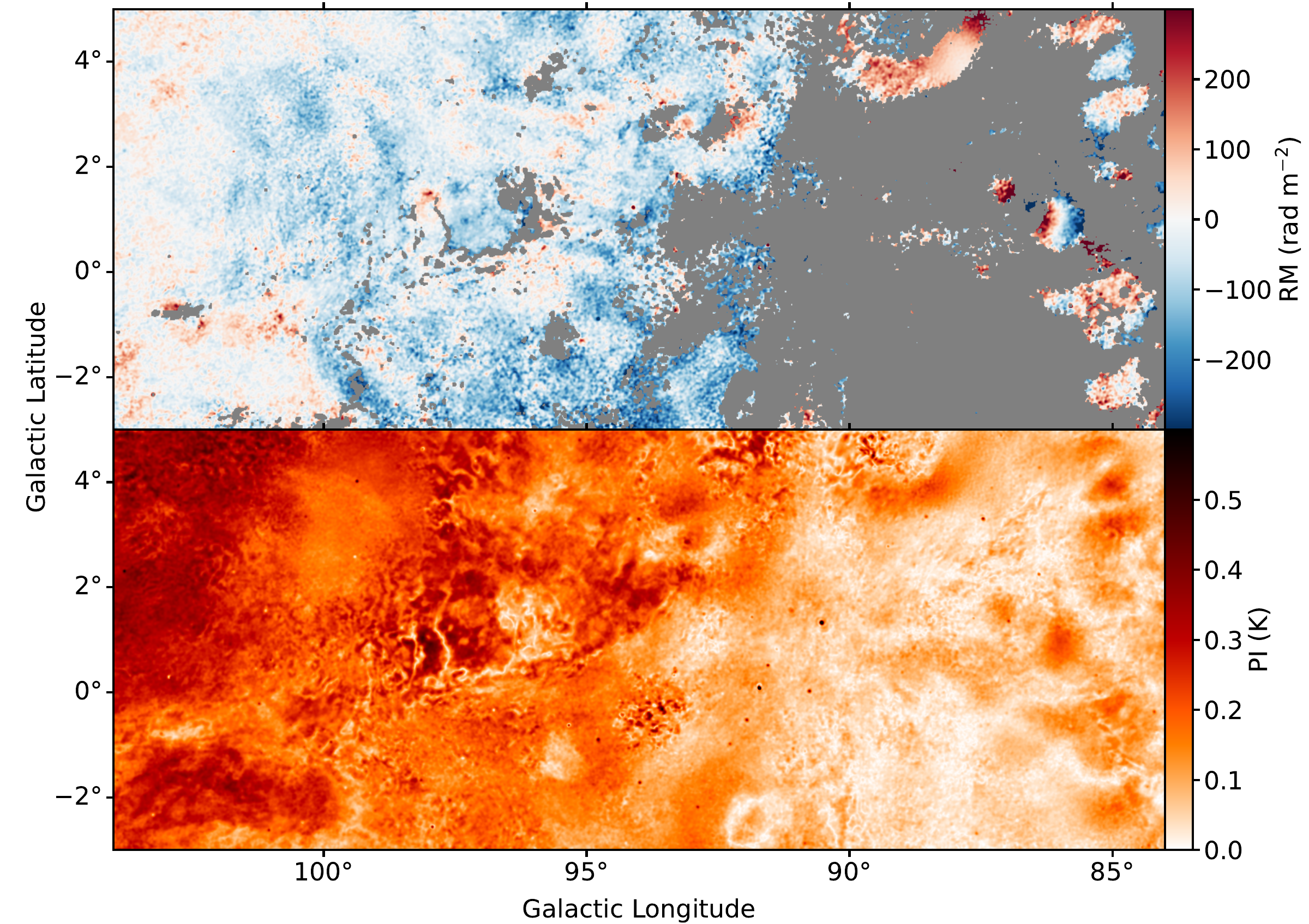} 
	\caption{Combined full-scale RM and PI maps for $84\arcdeg<\ell<104\arcdeg$, as in Figure~\ref{fig:fullmap1}. These maps cover the same region as the bottom panel of Figure~5 in L10.
 }
	\label{fig:fullmap2}
\end{figure*}

\begin{figure*}
	\centering
	\includegraphics[width=0.78\textwidth]{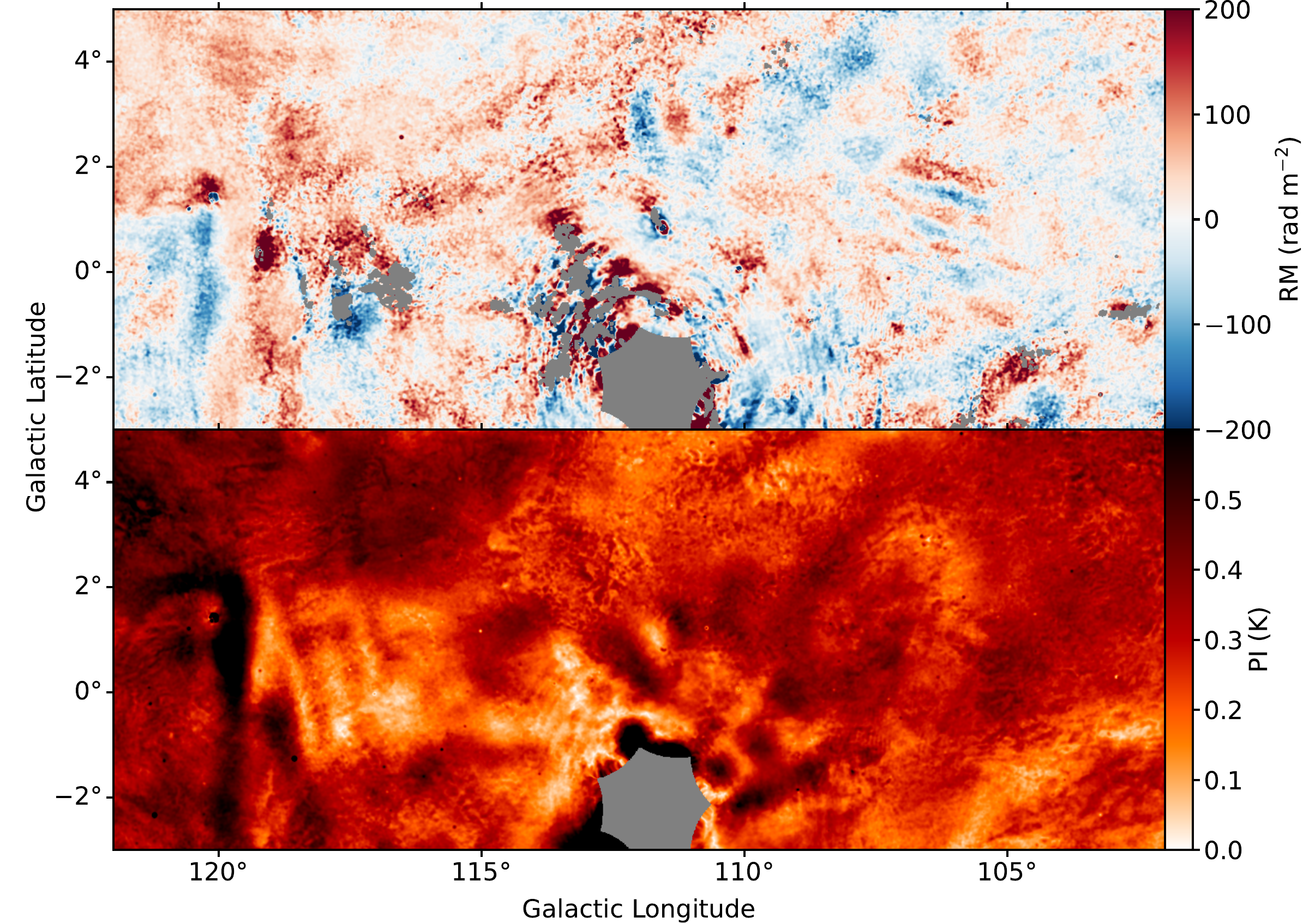} 
	\caption{Combined full-scale RM and PI maps for $102\arcdeg<\ell<122\arcdeg$, as in Figure~\ref{fig:fullmap1}. These maps cover the same region as the top panel of Figure~6 in L10. Note that the missing region at $\ell=112\arcdeg$ is due to masking around Cassiopeia A, which is too bright for the ST imaging. The hook-shaped feature in PI at $\ell=120\arcdeg$ is an artifact in GMIMS-HBN.
 }
	\label{fig:fullmap3}
\end{figure*}

\begin{figure*}
	\centering
	\includegraphics[width=0.78\textwidth]{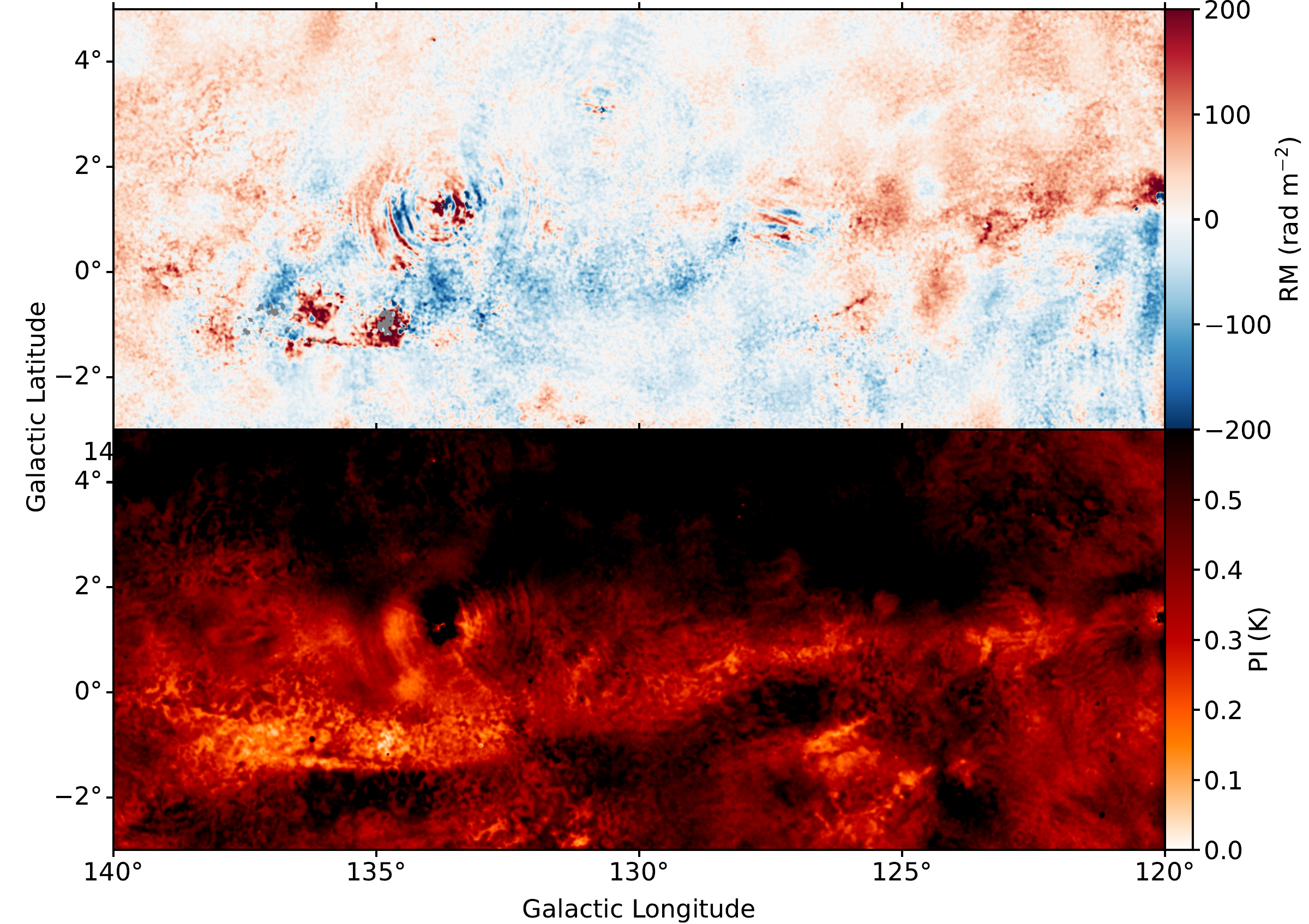} 
	\caption{Combined full-scale RM and PI maps for $120\arcdeg<\ell<140\arcdeg$, as in Figure~\ref{fig:fullmap1}. These maps cover the same region as the bottom panel of Figure~6 in L10.
 }
	\label{fig:fullmap4}
\end{figure*}

\begin{figure*}
	\centering
	\includegraphics[width=0.78\textwidth]{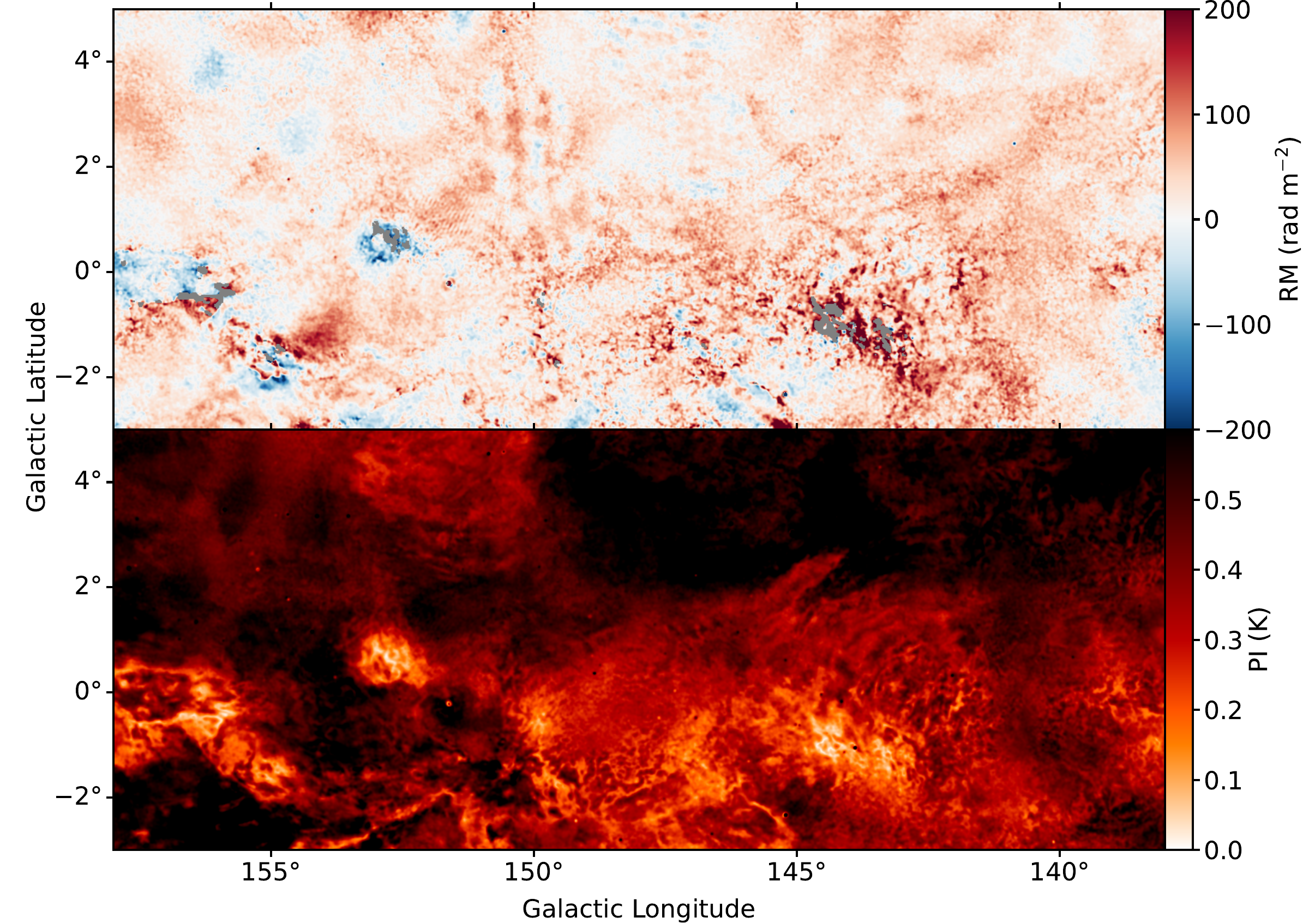} 
	\caption{Combined full-scale RM and PI maps for $138\arcdeg<\ell<158\arcdeg$, as in Figure~\ref{fig:fullmap1}. These maps cover the same region as the top panel of Figure~7 in L10.
 }
	\label{fig:fullmap5}
\end{figure*}

\begin{figure*}
	\centering
	\includegraphics[width=0.78\textwidth]{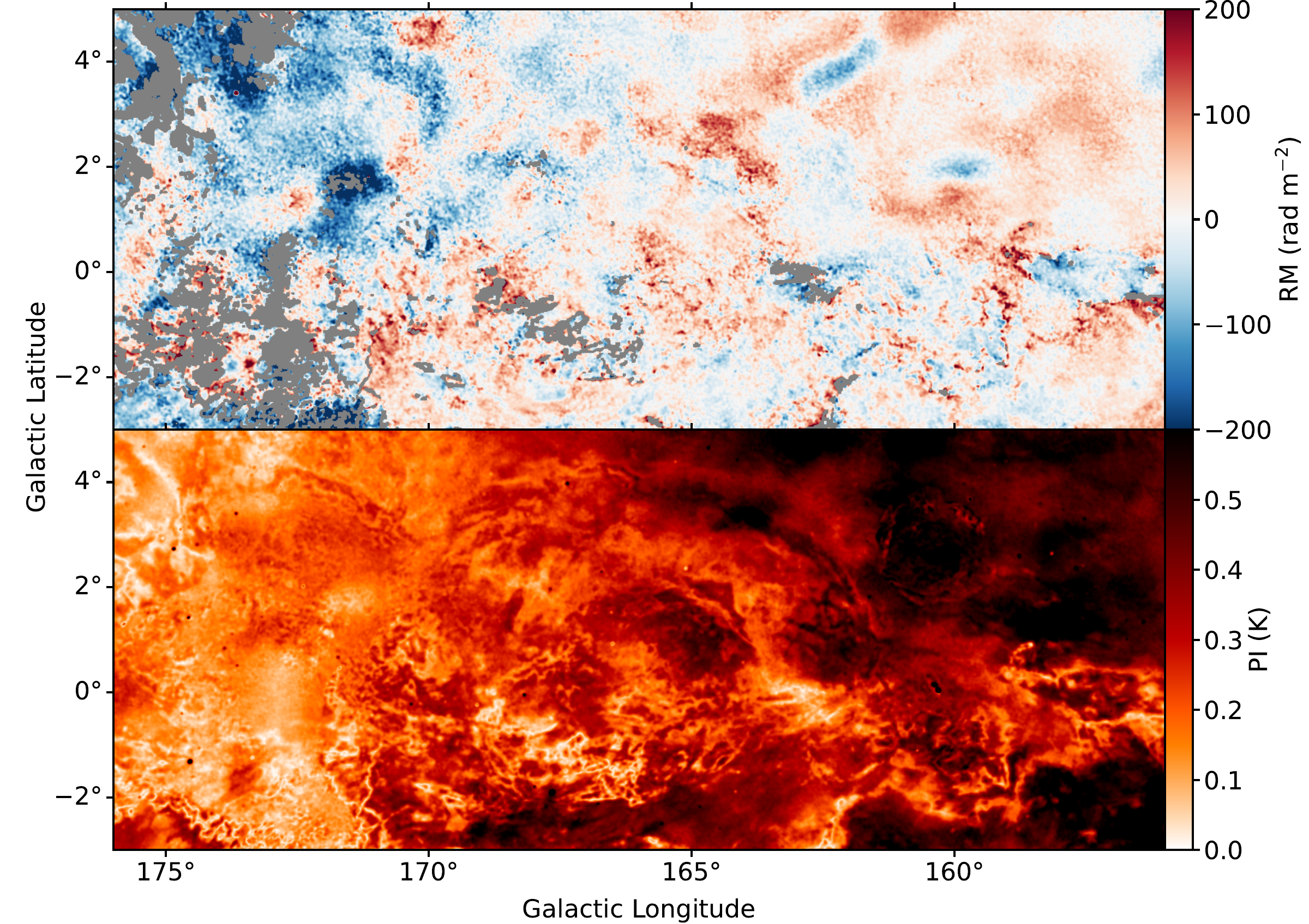} 
	\caption{Combined full-scale RM and PI maps for $156\arcdeg<\ell<176\arcdeg$, as in Figure~\ref{fig:fullmap1}. These maps include the region in the bottom panel of Figure~7 in L10, as well as previously unpublished CGPS Phase 3 data.
 }
	\label{fig:fullmap6}
\end{figure*}

\begin{figure*}
	\centering
	\includegraphics[width=0.78\textwidth]{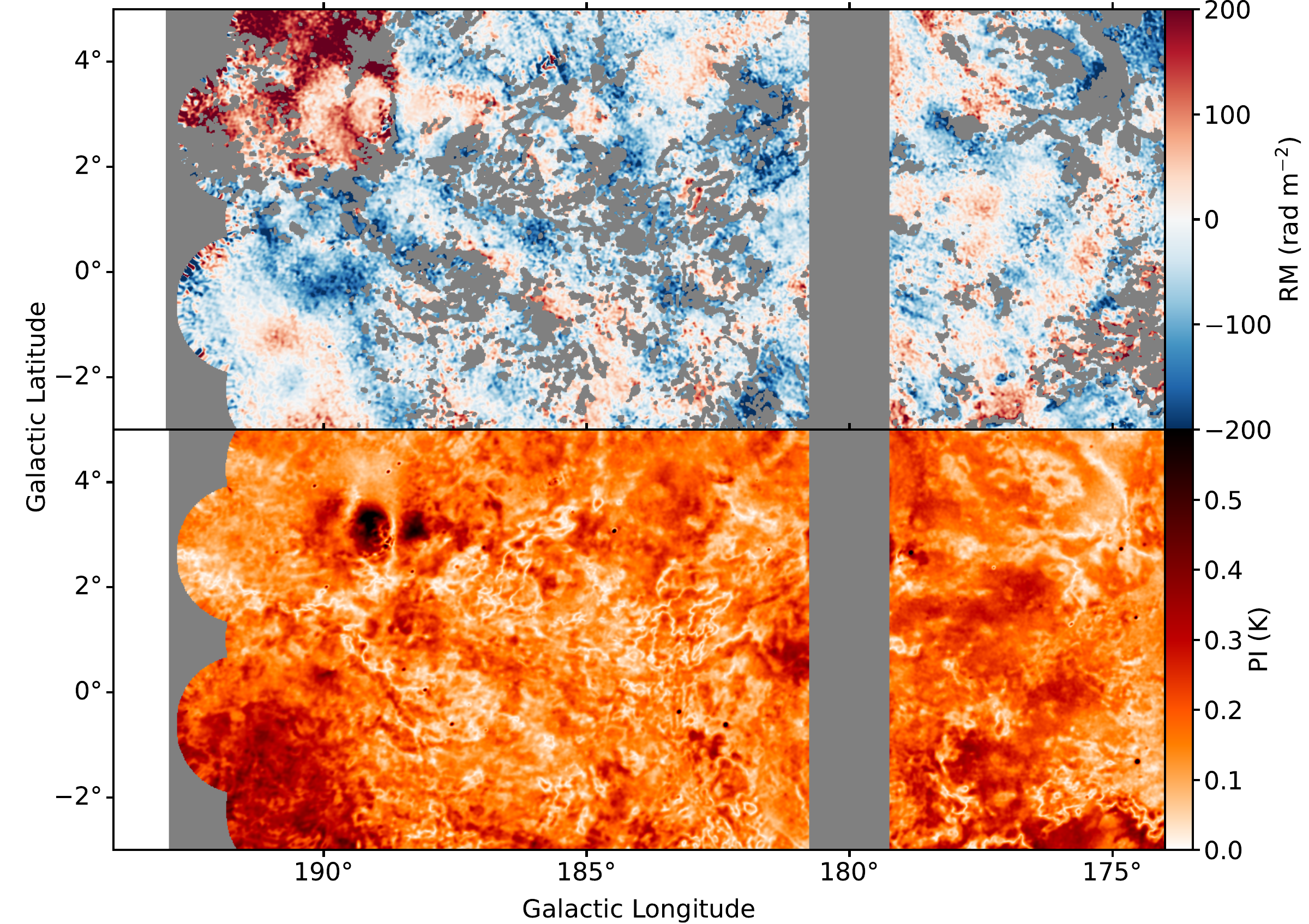} 
	\caption{Combined full-scale RM and PI maps for $174\arcdeg<\ell<194\arcdeg$, as in Figure~\ref{fig:fullmap1}. This region covers the high-longitude end of the previously unpublished CGPS Phase 3 data. The 1$\arcdeg$-wide strip at $\ell=180\arcdeg$ is masked owing to a discontinuity in the GMIMS-HBN data at this longitude producing artifacts in the Fourier transform steps described in Section~\ref{sec:steps_combine}.
 }
	\label{fig:fullmap7}
\end{figure*}

The PI maps shown in Figures~\ref{fig:fullmap1}-\ref{fig:fullmap7} are calculated as $\text{PI} = \sqrt{\langle Q \rangle^2+\langle U \rangle^2}$ where the averages $\langle Q \rangle$ and $\langle U \rangle$ are taken over the four ST frequency channels. As in L10, there is no polarization bias correction applied. The full-scale, combined PI maps reveal structures across a wide range of spatial scales, as shown also in L10. Over much of the longitude coverage, the small-scale structure in the polarized intensity is clearest where the surface brightness is lowest (PI$<0.3$~K), as seen, for example, near the midplane below the `Fan region' in Figures~\ref{fig:fullmap4}-\ref{fig:fullmap6} ($120\arcdeg$ to $160\arcdeg$). The differences between the PI maps shown here and those in L10 are primarily due to slight differences in the 26 m single-antenna maps produced for GMIMS-HBN compared to the W06 1.4 GHz map that was used as the zero-spacings in L10, and the lack of inclusion of the Effelsberg 100 m data in the present analysis. As discussed in Section~\ref{sec:discussion_gap}, the effect of the latter likely only leads to a few percent difference. A thorough comparison between the W06 1.4 GHz map and GMIMS-HBN is beyond the scope of this study, but we highlight some of the key differences here.

\begin{figure*}
	\centering
	\includegraphics[width=\textwidth]{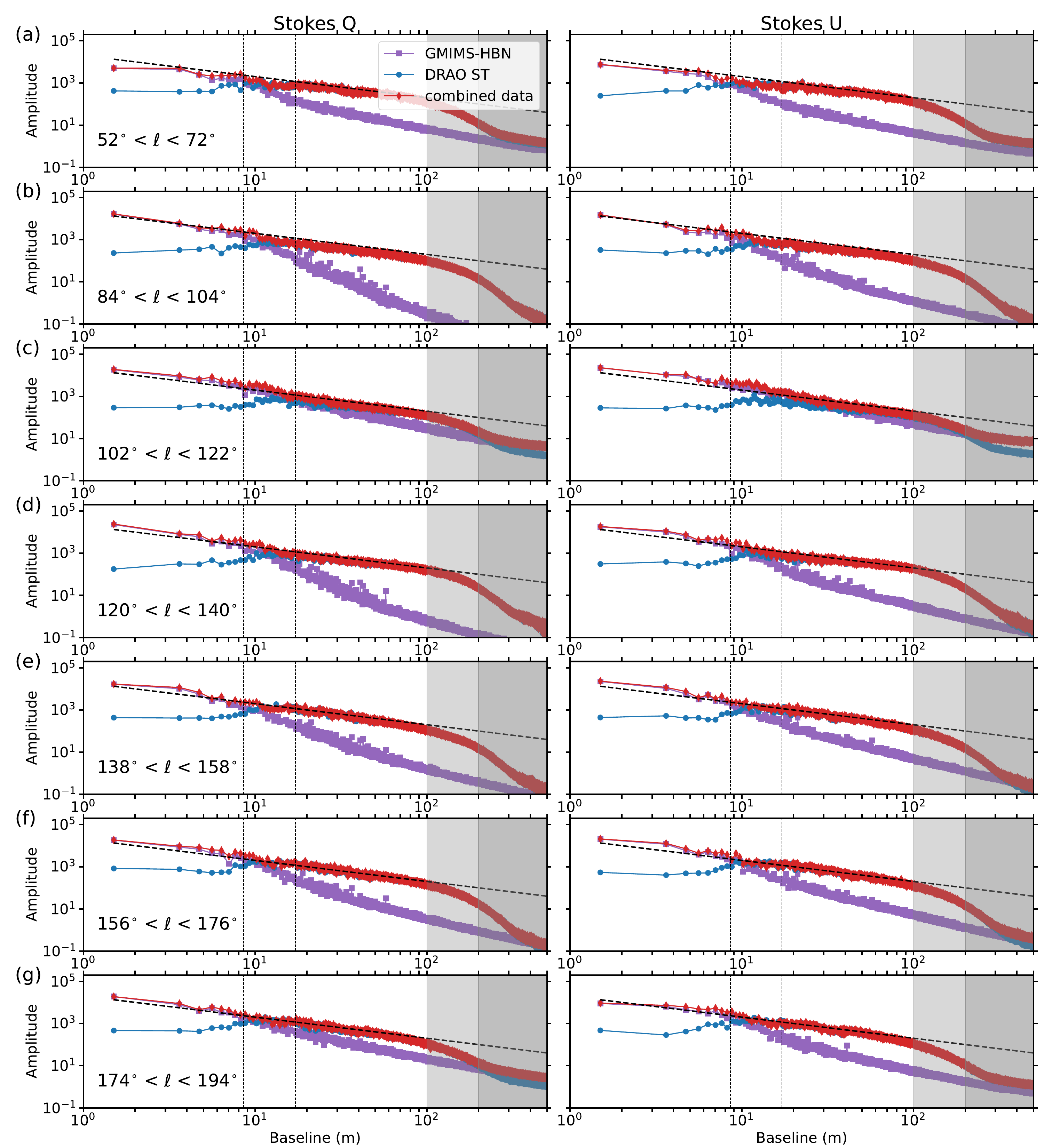} 
	\caption{Stokes $Q$ (left) and Stokes $U$ (right) power spectra for the GMIMS-HBN single-antenna (purple squares), the DRAO ST (blue circles) and the combined data (red diamonds). Rows (a)-(g) correspond to the regions shown in Figures~\ref{fig:fullmap1}-\ref{fig:fullmap7}, with the Cygnus X longitudes excluded from Figure~\ref{fig:fullmap1}. The light-gray regions indicate baselines between 100~m and 200~m, where the DRAO ST data convolved to $3'$ start to decline in sensitivity, dropping to 20\% at $\sim$200~m \citep[$\sim$600~m at the native $1'$ resolution;][]{Landecker_2000}. The dark-gray regions indicate baselines at which the convolved DRAO ST is no longer reliable. The dashed lines correspond to a power law with slope -1 (not a fitted line; intended only to facilitate comparison between the regions). Vertical dotted lines indicate the overlap region for the $uv$-domain feathering of the single-antenna and interferometric datasets.
 }
	\label{fig:powerspectra}
\end{figure*}

The W06 1.4 GHz survey was produced from drift scans with the telescope stationary on the meridian at fixed declinations in the range from $-29\arcdeg$ to $90\arcdeg$. In the time allocated for the survey, only 41.7\% of full Nyquist sampling was achieved in terms of spatial coverage. The \cite{Wolleben_2021} GMIMS-HBN survey was observed using declination scans up and down the meridian, covering declinations $-30\arcdeg$ to $87\arcdeg$, achieving 95\% of full Nyquist spatial sampling. The GMIMS-HBN polarization maps have fewer remaining scanning artifacts than the W06 1.4~GHz map. As an independent check, we compared both datasets with the 1411~MHz Dwingeloo polarization maps from \cite{Spoelstra_1976}, finding a stronger correlation between GMIMS-HBN and Dwingeloo than between W06 and Dwingeloo. Comparisons of GMIMS-HBN and W06 to Dwingeloo are presented in \cite{Wolleben_2021} and \cite{Wolleben_2006} respectively.

The most significant difference in large-scale PI structure between the present study and L10 is in the Fan region ($120\arcdeg<\ell<160\arcdeg$). Comparing Figure~\ref{fig:fullmap4} to the bottom panel of Figure~6 in L10, the amount of polarized signal is higher for $b >2\arcdeg$ in our combined full-scale maps than in the L10 CGPS map. Given its improved spatial sampling, using GMIMS-HBN as the single-antenna contribution likely gives a more accurate representation of the highly polarized Fan region than using the W06 1.4 GHz map. This is supported by the W06 comparison with Dwingeloo, revealing an underestimate of PI in this region in W06. 

In our combined full-scale PI map, the Cygnus X region ($77\arcdeg<\ell<84\arcdeg$) and Cygnus A ($\ell=78\arcdeg$, $b=5 \arcdeg$; Figure~\ref{fig:fullmap1}) are heavily contaminated by Stokes~$I$ leakage into polarization and therefore should not be trusted. This effect around Cygnus X is less present in the W06 1.4 GHz map compared to GMIMS-HBN, which may be due to undersampling and smoothing of the scans in this region in W06. Cygnus A leakage appears in both W06 and GMIMS-HBN, but its effect appears to be reduced by the inclusion of the Effelsberg 100 m data in the CGPS combined map. Although our simulated, mock observations (Section~\ref{sec:discussion_simulate}) indicate that the missing spacings do not have a strong effect on the combined maps, those results are likely not valid in regions surrounding bright sources that produce residual polarization artifacts even after leakage correction.

Finally, the hook-shaped feature in PI at $\ell=120\arcdeg$, spanning $\sim5\arcdeg$ in latitude (Figure~\ref{fig:fullmap3}) results from a vertically extended artifact present in GMIMS-HBN, interacting with the bottom edge of the bright polarized emission in the Fan region. Apart from these discrepancies, our full-scale PI maps are very similar to those in L10 despite the exclusion of the Effelsberg 100 m data from the current study and the slight differences between the GMIMS-HBN and W06 maps.

We show power spectra for the GMIMS-HBN, DRAO ST, and combined full-scale Stokes $Q$ and $U$ maps in Figure~\ref{fig:powerspectra}, in panels corresponding to Figures~\ref{fig:fullmap1}-\ref{fig:fullmap7}, with the Cygnus X region excluded from Figure~\ref{fig:fullmap1}. An in-depth power spectrum analysis is beyond the scope of this paper, with details for the CGPS already presented in \cite{Stutz_2014}, and here we highlight a few key features in the present dataset. First, for most longitude ranges, the DRAO ST and GMIMS-HBN data intersect as expected in the overlap region of the $uv$-plane (between the dashed lines in Figure~\ref{fig:powerspectra}), with no substantial noise or deviations from the overall trend in either dataset. This confirms the negligible impact of the small $uv$-coverage gap. Second, the DRAO ST data dominate the combined full-scale power beyond the outer boundary of the feathered region ($\sim 18$~m), as expected. Finally, the GMIMS-HBN data dominate the power at the shortest baselines, although there is a longitude-dependent difference between GMIMS-HBN and the DRAO ST data. At the highest and lowest longitudes (corresponding to the lowest declinations in the CGPS survey) foreshortening of the DRAO ST baselines yields effective sensitivity to larger scales. This is likely the cause of the smaller difference between GMIMS-HBN and the DRAO ST power near the shortest ST baselines at high and low longitudes (Figure~\ref{fig:powerspectra} panels (a) and (g)) compared to intermediate longitudes. The exception to the patterns described above is the $102\arcdeg<\ell<122\arcdeg$ range (Figure~\ref{fig:powerspectra} panel c), in which instrumental polarization around Cassiopeia A contaminates the GMIMS-HBN data.

\subsection{Rotation measure maps}\label{sec:results_RMmaps}
\begin{figure*}
	\centering
	\includegraphics[width=0.8\textwidth]{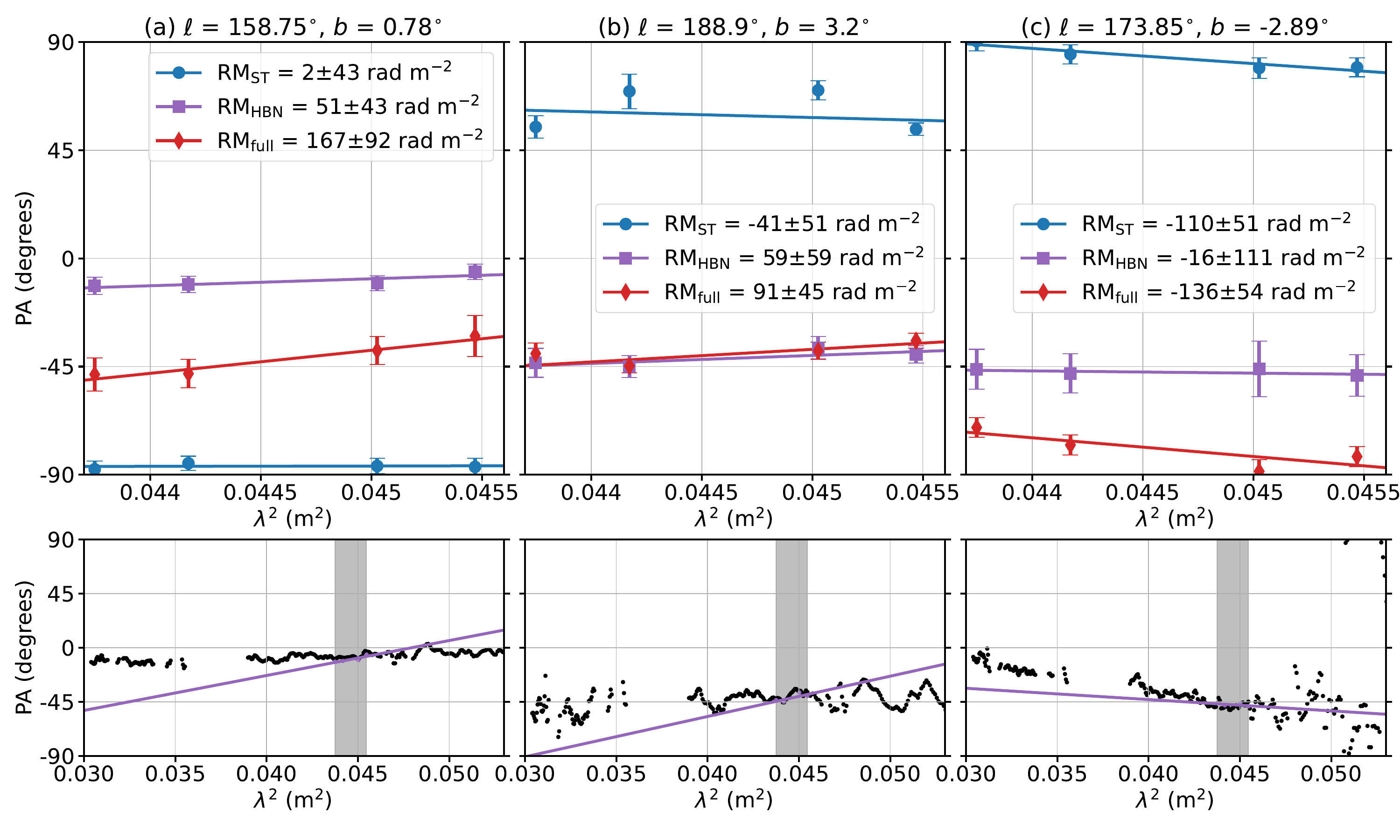} 
	\caption{Sample linear fit RM values for three lines of sight, comparing the single-antenna-only data (GMIMS-HBN; purple, denoted `HBN'), aperture-synthesis-only data (DRAO ST; blue, denoted `ST'), and the combined full-scale data (red, denoted `full'). The bottom panels show the full-band GMIMS-HBN along with the linear fit to GMIMS-HBN for the four-channel 35 MHz bandwidth of the DRAO ST (indicated by the shaded gray $\lambda^2$ range). The three lines of sight shown here in panels (a)-(c) are indicated by crosses in Figures~\ref{fig:sh216}, \ref{fig:ic443}, \ref{fig:bowtie}.
	}
	\label{fig:linear_fits}
\end{figure*}

The RM maps shown in Figures~\ref{fig:fullmap1}-\ref{fig:fullmap7} are calculated by determining a linear fit to the polarization angle $\tau=0.5\tan^{-1}{U/Q}$ as a function of $\lambda^2$ across the four CGPS bands for each pixel in the map. Although we tested applying RM synthesis to these four channels, this does not provide a significant advantage over a linear fit owing to the narrow $\lambda^2$ coverage (and corresponding broad RMSF), apart from a slight reduction in bandwidth depolarization in the case of high RM values.  Examples of the linear fit compared between GMIMS-HBN data alone, the DRAO ST data alone, and the combined full-scale data are shown in Figure~\ref{fig:linear_fits}. 

With the 35~MHz bandwidth centered on 1.4~GHz, we are unable to directly detect Faraday complexity in the mapped region. An alternative approach is to examine the change in PI across the band compared to the expected change due to the spectral index alone. We find that for almost $90\%$ of the map the PI changes by more than a spectral index of -2.7 would cause, and in roughly half of those pixels the slope has the opposite sign (lower PI for lower frequencies). This may indicate the presence of some degree of Faraday complexity across much of the map, but it is impossible to further constrain the nature of it with the limited bandwidth. The presence of Faraday complexity does not necessarily invalidate the interpretation of the RMs calculated over the 35~MHz bandwidth. If the complexity takes the form of a Burn slab, for example, the polarization angle as a function of wavelength squared will still be linear, and the RM derived across a narrow band will represent half of the extent of the slab in Faraday depth. Scenarios that deviate from a Burn slab, such as those highlighted in Figure 7 of \cite{Ordog_2019}, can indeed produce RM values over a narrow band that are difficult to interpret, and RM values we derive should therefore be interpreted with caution, in conjunction with analysis of the full GMIMS-HBN 1280-1750~MHz data (see Section~\ref{sec:discussion_polhorizon} for further discussion).

In the RM maps, we mask regions of low PI. We derived an average of 0.02~K as the noise in the full-scale PI map from measurements of the noise in the individual-channel Stokes $Q$ and $U$ maps. Setting a 5$\sigma$ minimum for signal-to-noise ratio (S/N) gives a PI threshold of 0.1~K, which we apply to both the PI and RM maps. The noise in the DRAO ST-only data is similar, and we use the same 0.1~K PI threshold for those maps. This results in slightly more masking of the maps compared to \cite{Ordog_2017}, in which a lower estimate for the noise in the Stokes $Q$ and $U$ maps was used.

Although \cite{Ordog_2019} showed that it was possible to observe large-scale RM patterns with aperture-synthesis-only data by binning in Galactic longitude across the CGPS range, the inclusion of the single-antenna data now allows for spatially coherent structures to be observed in the resulting RM map. The largest-scale structures observed in the full-scale RM maps in Figures~\ref{fig:fullmap1}-\ref{fig:fullmap7} correspond to structures seen in the equivalent 35~MHz bandwidth GMIMS-HBN data RM maps. This is evident when binning the single-antenna-only and full-scale RMs by longitude, which we discuss in Section~\ref{sec:discussion_polhorizon}. However, we see plenty of small-scale structures in the RM maps as well, which are traced by the aperture-synthesis component. In some cases, the addition of the single-antenna component to the aperture-synthesis data has a sufficiently large effect to produce changes in the sign of the RM. The impact of this is important when interpreting aperture-synthesis or single-antenna RM data alone. We explore a few interesting regions in Section~\ref{sec:discussion_uses} in terms of the separate and combined data RMs.

There are a few regions where artifacts appear in the combined full-scale RM map. The instrumental polarization leakage in the Cygnus X region, described in Section~\ref{sec:results_PImaps}, leads to patchy, rapidly changing, large-magnitude RM values in this region. These almost certainly do not represent true RM values associated with these lines of sight. In regions around other bright sources, such as Cassiopeia A ($\ell=112\arcdeg$, $b=-2\arcdeg$) and W3 ($\ell=134\arcdeg$, $b=1\arcdeg$), there are instrumental polarization effects arising from complex gain errors in the DRAO ST data, which lead to ring-shaped patterns in the RM maps around these sources. We must be careful not to interpret these as true RM gradients.

\section{Discussion}\label{sec:discussion}
In this section we discuss the roles of single-antenna and aperture-synthesis data in the RM maps, illustrate the use-cases with some examples from the combined dataset described above, and describe the effects of beam depolarization and bandwidth limitations.

\subsection{The roles of single-antenna and aperture-synthesis data}\label{sec:roles}
In the case of total intensity (Stokes $I$) or polarized intensity, single-antenna data are always necessary in addition to aperture-synthesis data for recovering the flux on the largest spatial scales. For RM maps, aperture-synthesis data alone can be sensitive to large-scale patterns in the cases where there is sufficient structure on small scales in Stokes $Q$ and $U$ for there to be a polarized signal detectable by interferometry, as shown in \cite{Gaensler_2001, Haverkorn_2003a, Haverkorn_2003b, Ordog_2017}. In these cases aperture-synthesis-only RM maps may largely resemble their single-antenna counterparts. However, single-antenna data are needed when the diffuse polarized emission is too spatially smooth to be detectable by aperture-synthesis data alone. This is the case over much of the intermediate longitudes in the CGPS (approximately $\ell=70\arcdeg$ to $\ell=160\arcdeg$). 

\begin{figure*}
	\centering
	\includegraphics[width=1\textwidth]{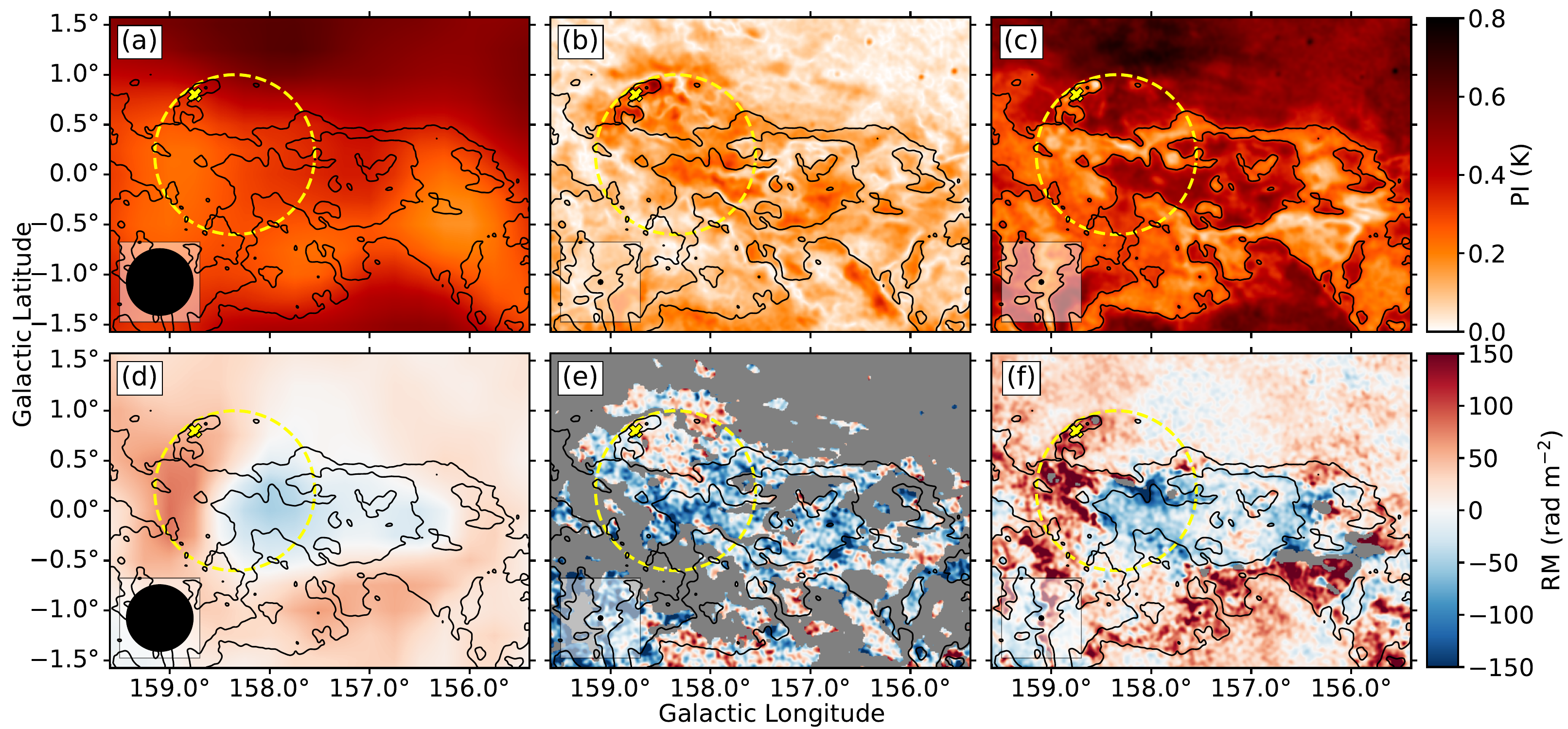} 
	\caption{The GMIMS-HBN, DRAO ST, and full-scale data in the Sh2-216 PN region. Top row (a-c): PI calculated as $PI = \sqrt{\langle Q \rangle^2+\langle U \rangle^2}$. Bottom row (d-f): RM for pixels with PI$>0.1$~K. Left column (a,d): GMIMS-HBN single-antenna-only data. Middle column (b,e): DRAO ST aperture-synthesis-only data. Right column (c,f): combined full-scale data. Black circles in the lower left corners indicate the beam size ($40'$ for the single-antenna, $\sim3'$ for the aperture-synthesis and full-scale maps). The yellow cross indicates the LOS shown in Figure~\ref{fig:linear_fits}(a). Black contours indicate PI$=0.3$~K on the full-scale map (c) and are shown in all panels for ease of comparison. The yellow dashed circle corresponds to the disk of the PN as determined in R08.
	}
	\label{fig:sh216}
\end{figure*}

\begin{figure}
	\centering
	\includegraphics[width=\hsize]{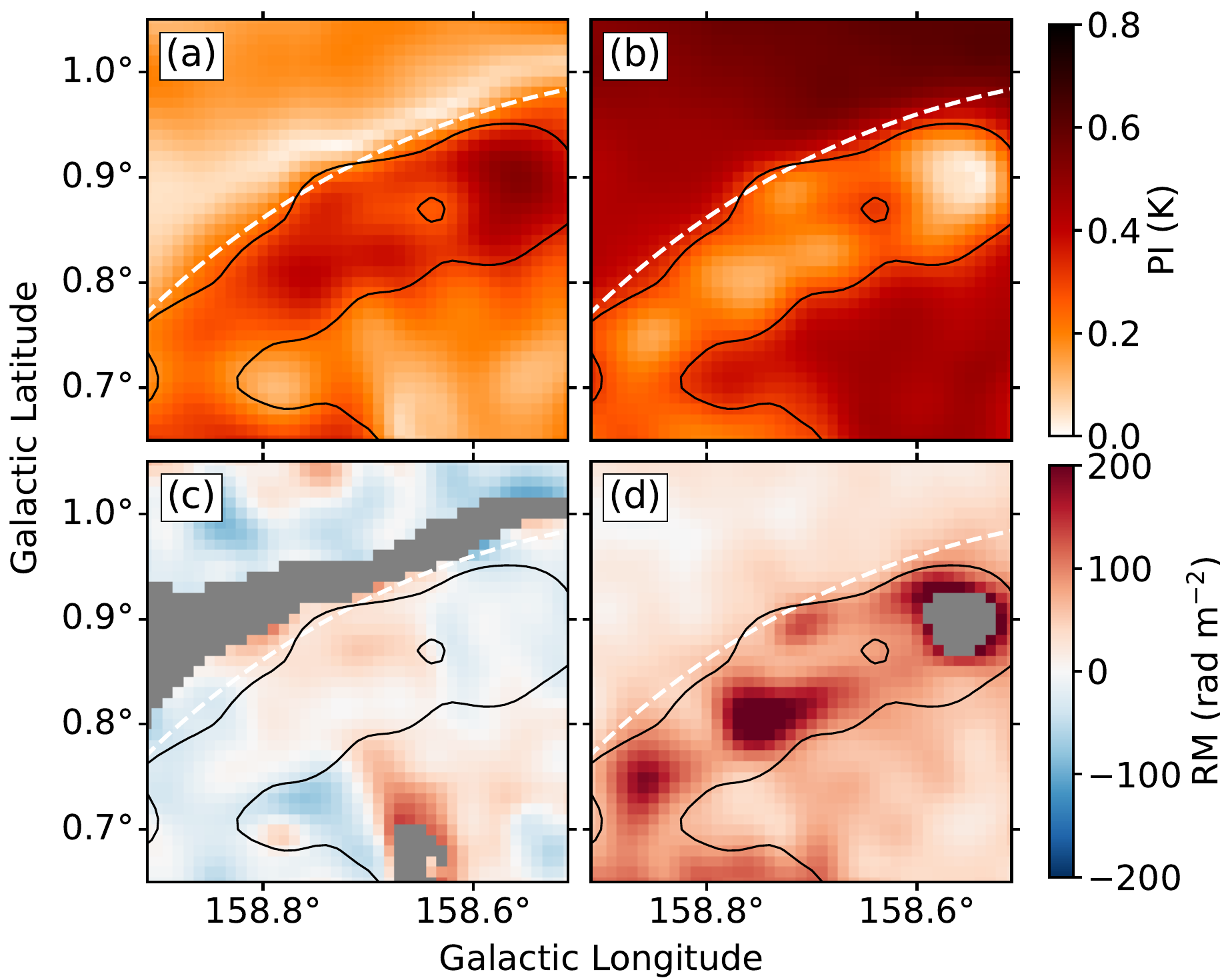} 
	\caption{The northeast arc of Sh2-216 showing the `polarization knots' studied by \citet[R08]{Ransom_2008} and approximately matching the region shown in their Figure~3. As in Figure~\ref{fig:sh216} the top row shows PI and the bottom row shows RMs, for DRAO ST data (panels a, c) and the combined dataset (panels b, d). The white dashed arc corresponds to the disk of the planetary nebula as determined in R08 (also shown as a yellow circle in Figure~\ref{fig:sh216}). Note that the resolution is degraded to $3'$ compared to the $1'$ CGPS resolution in R08.
 }
	\label{fig:sh216zoom}
\end{figure}

In \cite{Ordog_2019} the masking criteria for the CGPS ST maps consisted of removing pixels with high Stokes $I$ to avoid discrete objects such as supernova remnants (SNRs), setting a minimum S/N threshold, and excluding pixels with low probability of a linear fit in the RM calculation. Among these masking criteria, the S/N thresholding led to the most substantial reduction in the remaining pixels included in the analysis. While the low pixel count was still sufficient to trace the overall trend in diffuse emission RMs across Galactic longitude by binning the values in longitude \citep[Figure~4 in ][Figure~\ref{fig:RMversuslong}a in this paper]{Ordog_2019}, it is prohibitive for producing useful diffuse RM maps in the low-S/N regions. By contrast, inclusion of the single-antenna data allows for high-S/N RM maps across nearly all CGPS Galactic longitudes. As in the case of total power and PI maps, if there is also small-scale structure in the spatial distribution of Faraday depths (i.e., small-scale magnetic field or electron density fluctuations), then aperture-synthesis data contribute valuable high-resolution information.

\subsection{Applications of full-scale RM analysis}\label{sec:discussion_uses}
Here we highlight examples from the combined full-scale maps in comparison with the single-antenna-only and aperture-synthesis-only maps to demonstrate the importance of the inclusion of each of these components in terms of tracing RM structures across spatial scales. Although much of the CGPS region, particularly for $\ell > 92\arcdeg$, has smooth structure in RM, indicating that single-antenna data alone might be sufficient, there are scenarios in which small-scale structures are traced by the aperture-synthesis data but with insufficient S/N to make aperture-synthesis-only mapping effective. In such cases, the addition of the single-antenna data appears to provide the necessary sensitivity, while the aperture-synthesis data provide the spatial details.

\subsubsection{Planetary Nebula Sh2-216}\label{sec:discussion_sh216}
The polarization structure of the planetary nebula (PN) Sh2-216 (Figure~\ref{fig:sh216}) has been previously studied by \citet[][hereafter R08]{Ransom_2008} using the band-averaged CGPS data from L10. In R08 the authors noted the presence of the northeastern arc of Sh2-216 (Figure~\ref{fig:sh216zoom}) in maps of PI and polarization angle, along with bands of low polarized intensity and changes in polarization angle stretching east-west across the central and southern regions of the PN that were not confirmed to be directly associated with Sh2-216. Based on the changes in polarization angle across a series of `polarization knots' distributed along the northeastern arc, R08 derived an estimated RM of $-43\pm10\radmsq$ in this region.

The northeastern arc of Sh2-216 is also apparent as a depolarized region in our combined full-scale PI map (Figure~\ref{fig:sh216}c, \ref{fig:sh216zoom}b), with the `polarization knots' appearing as small patches of enhanced brightness in the DRAO ST-only PI map and enhanced depolarization in the combined full-scale PI map. However, in our full-scale RM map (Figure~\ref{fig:sh216}f, \ref{fig:sh216zoom}d) the arc stands out as having strongly \textit{positive} RM values (RM$>100 \radmsq$). R08 noted being unable to determine RMs along the arc from the DRAO ST-only four-band data, and we find that the arc does not stand out from the surroundings in our ST-only RM map (Figure~\ref{fig:sh216}e, \ref{fig:sh216zoom}c). The inclusion of the smooth, predominantly positive RMs from the GMIMS-HBN data (Figure~\ref{fig:sh216}d) results in the `polarization knots' standing out with enhanced positive RM along the arc. 

The change in polarization angle from outside the `polarization knots' moving toward their centers that we observe in a single channel of our full-scale data agrees with the findings of \cite{Ransom_2008} (a negative change in angle, corresponding to clockwise rotation and negative RM). The discrepancy between this result and the RMs derived from the linear fits to our combined full-scale four-channel data may be due to the full-scale data being dominated by large-scale RM structures. Using the spatial variation in polarization angle across the arc may probe smaller-scale RM features than the overall RM of the region. The best practice is likely to examine both single-antenna and aperture-synthesis data in addition to full-scale data in order to thoroughly study Faraday rotation of structures on different scales.

It is important to note that, in general, uncertainties in the RM values can also result from Faraday complexity not observable in the 35~MHz four-band CGPS frequency range. There may be Faraday complexity in this region arising from Sh2-216 breaking the diffuse emission into background and foreground components, with differing Faraday rotation, and thus the RM values in this region should be interpreted with caution. The bottom panels of Figure~\ref{fig:linear_fits} show the polarization angle as a function of $\lambda^2$ for the GMIMS-HBN 1280-1750~MHz coverage (not available for the DRAO ST data). In the bottom panel of Figure~\ref{fig:linear_fits}a, the overall slope of the GMIMS-HBN polarization angle versus $\lambda^2$ does not agree with the slope over the 35~MHz bandwidth, meaning that the narrower frequency range is picking up a localized trend in the broader spectrum. Although in this case the broader GMIMS-HBN spectrum does not indicate Faraday complexity, the addition of an interferometric component could also change the spectral profile. This is a limitation of the current experiment and highlights the importance of \textit{broadband} diffuse emission interferometric data in future surveys.

While the region of reduced PI to the southwest of the PN ($\ell\sim158\arcdeg$, $b\sim0\arcdeg$) shows hints of predominantly negative RM values in the aperture-synthesis data alone, it is transformed into a coherently negative RM structure in contrast to its positive RM surroundings with the inclusion of the single-antenna data. For the Sh2-216 region, although the aperture-synthesis data alone reveal PI structure along the edge of the PN, the single-antenna data are necessary to recover the coherent Faraday rotation structure. Wider frequency coverage for the aperture-synthesis data will help anchor the magnitudes of the RMs across the small-scale structures and may also reveal Faraday complexity along these lines of sight.

\subsubsection{Region near Supernova Remnant IC 443}\label{sec:discussion_ic443}
\begin{figure*}
	\centering
	\includegraphics[width=1\textwidth]{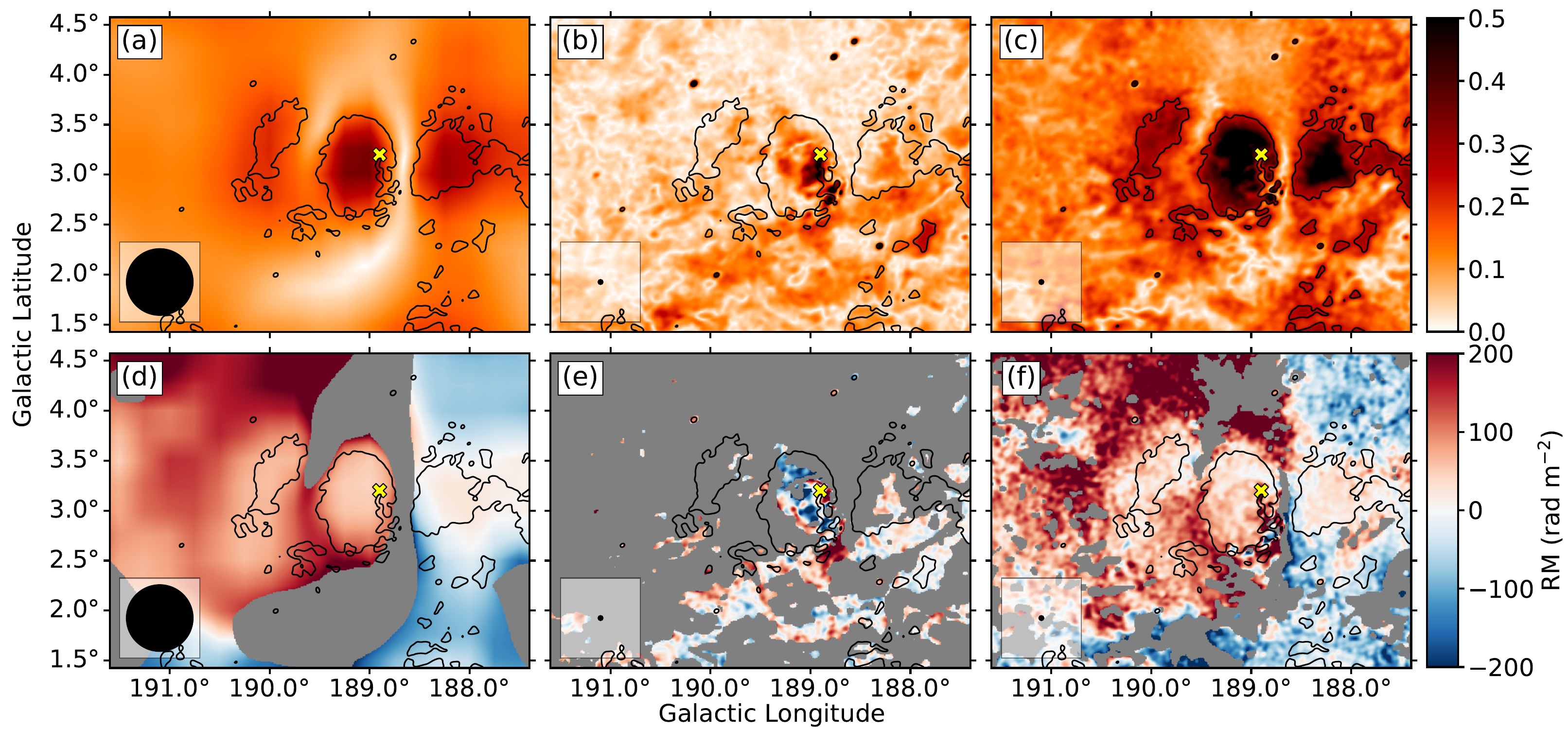} 
	\caption{The GMIMS-HBN, DRAO ST, and full-scale data in the IC 443 SNR region. Panels are the same as in Figure~\ref{fig:sh216}. The yellow cross indicates the LOS shown in Figure~\ref{fig:linear_fits}(b). Black contours indicate PI$=0.25$~K on the full-scale map (c).
	}
	\label{fig:ic443}
\end{figure*}

\begin{figure}
	\centering
	\includegraphics[width=0.9\hsize]{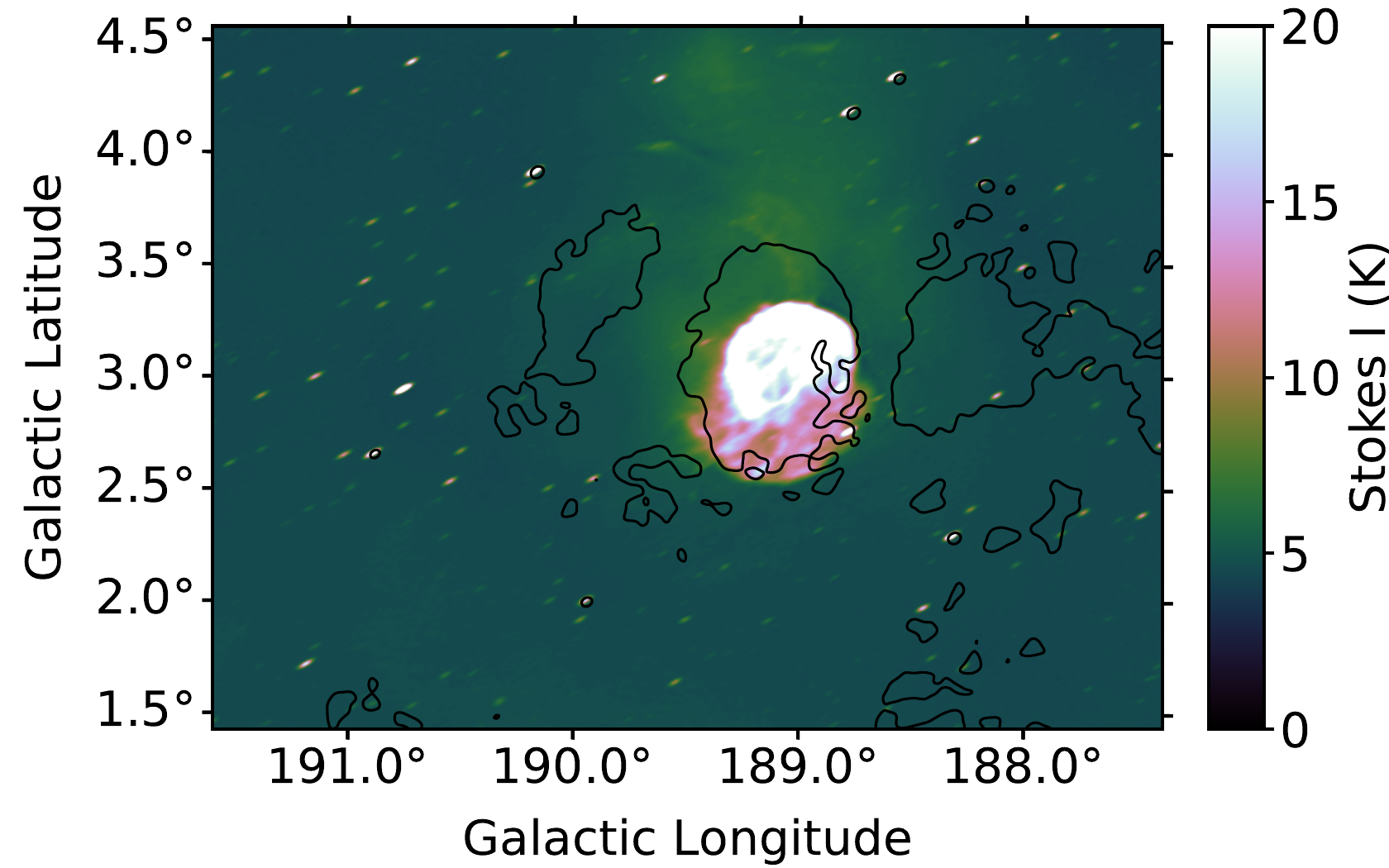} 
	\caption{CGPS Stokes $I$ in the IC 443 supernova remnant region. Black contours indicate PI$=0.25$~K on the full-scale map.
	}
	\label{fig:ic443StokesI}
\end{figure}

At the high-longitude end of the CGPS range, there is a sharp gradient in the RMs, transitioning from negative to positive values near $\ell=189\arcdeg$ at $b > 2\arcdeg$ (Figure~\ref{fig:fullmap7}). This gradient occurs across the southwestern edge of the SNR IC~443 (Figure~\ref{fig:ic443}). The gradient also correlates with a decrease in PI in the aperture-synthesis-only data from southwest to northeast. This lack of interferometric polarized emission in the northeastern corner of Figure~\ref{fig:ic443}, which includes a large portion of the SNR itself, is likely due to the polarized emission being too spatially smooth for the ST to detect. This is purely a polarization and Faraday rotation effect; there is no corresponding gradient or change in the scales of structures in total intensity, as seen in Figure~\ref{fig:ic443StokesI}. 

IC~443 being a relatively bright object compared to the surrounding diffuse emission may make it subject to residual polarization leakage. However, the angular size of the SNR in total intensity ($\sim1.5$\arcdeg$\times2$\arcdeg) makes it resolved by the 40$^{\prime}$ beam of GMIMS-HBN. We observe $\sim5\%$ polarization in IC~443, which is significant compared to the residual instrumental polarization of $0.3\%$ for beam-filling structures estimated in \cite{Wolleben_2021}. We therefore conclude that we can trust the polarized emission we observe on the SNR itself, not only in the surrounding region. The inclusion of the single-antenna polarization data reveals the positive RM values inside the SNR and in the northeastern region. The small-scale structures that appear as negative RMs in the aperture-synthesis-alone data become slight fluctuations on the overall positive RM of the SNR, as illustrated in the example LOS shown in Figure~\ref{fig:linear_fits}b. As for the case of Sh2-216, the full-band GMIMS-HBN traces an overall shallower slope of polarization angle versus $\lambda^2$, but with hints of Faraday complexity (bottom panel, Figure~\ref{fig:linear_fits}b).

\subsubsection{$\ell=173\arcdeg$ H II Complex}\label{sec:discussion_173}
\begin{figure*}
	\centering
	\includegraphics[width=1\textwidth]{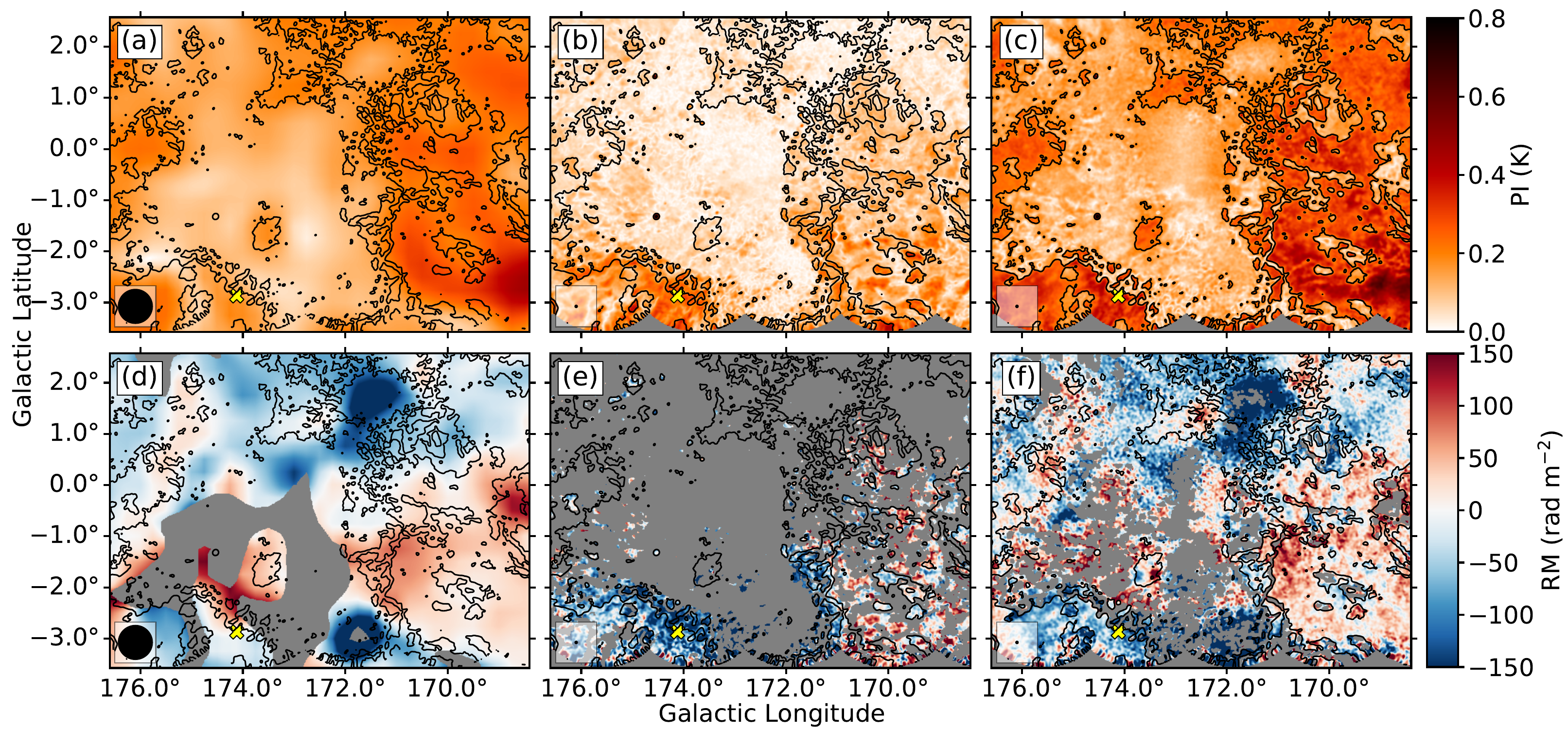} 
	\caption{The GMIMS-HBN, DRAO ST, and full-scale data in the $\ell=173\arcdeg$ H II complex. Panels are the same as in Figure~\ref{fig:sh216}. The yellow cross indicates the LOS shown in Figure~\ref{fig:linear_fits}(c). Black contours indicate PI$=0.2$~K on the full-scale map (c).
	}
	\label{fig:bowtie}
\end{figure*}

\begin{figure*}
	\centering
	\includegraphics[width=0.8\textwidth]{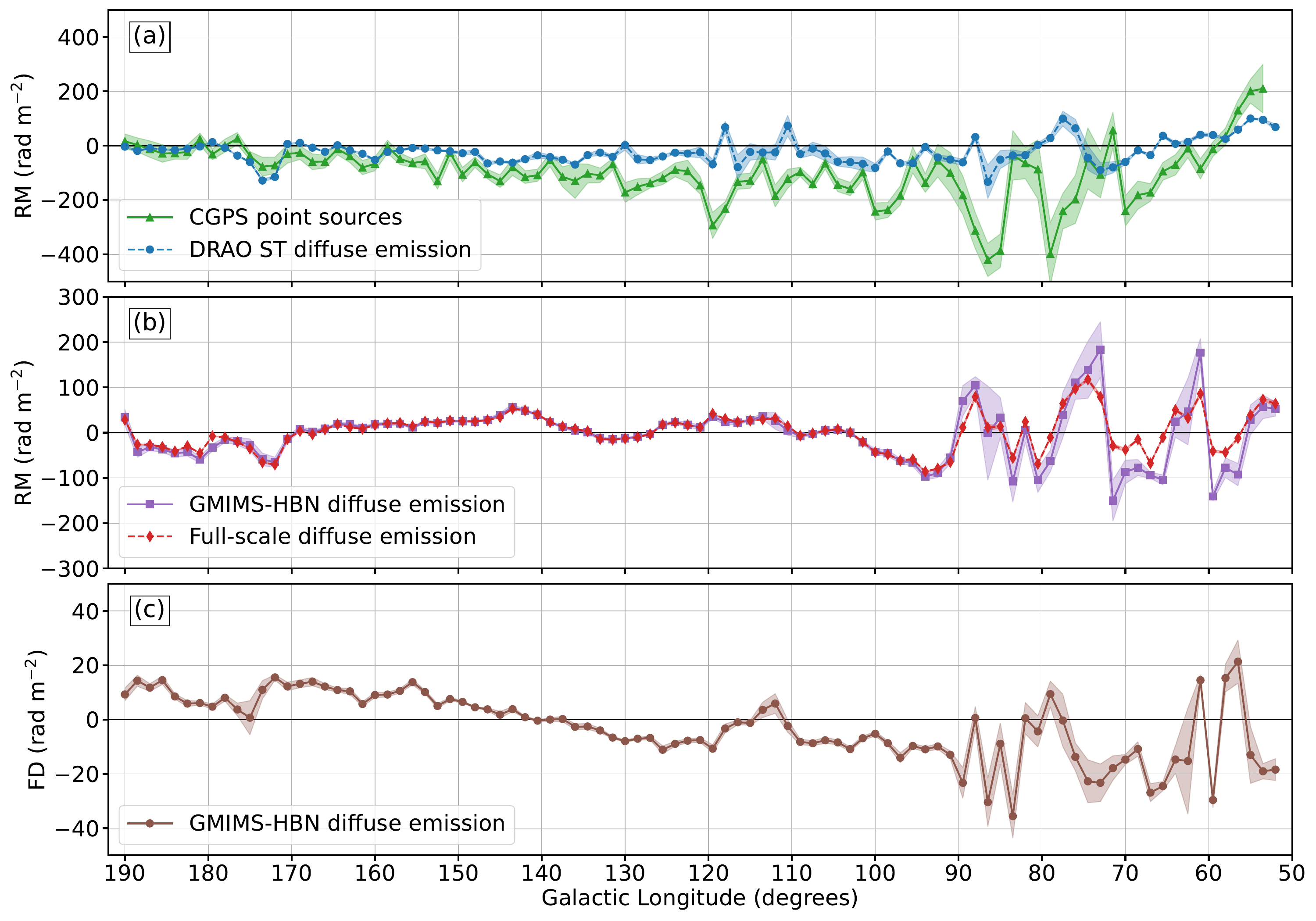} 
	\caption{Rotation Measures and peak Faraday depths as a function of Galactic longitude in the latitude range $-3.5\arcdeg\leq b \leq 5.5\arcdeg$ for diffuse synchrotron emission compared to compact polarized point sources. (a) aperture-synthesis-only data:  extragalactic point source RMs from the CGPS and (green), and the surrounding extended diffuse emission RMs (DRAO ST; blue). (b) single-antenna-only diffuse emission RMs (GMIMS-HBN; purple) and full-scale combined diffuse emission RMs (red) both using the CGPS 4-channel, 35 MHz bandwidth as in Figures~\ref{fig:fullmap1}-\ref{fig:fullmap7}. (c) single-antenna-only diffuse emission peak Faraday depth from RM Synthesis applied to the full GMIMS-HBN 1280-1750 MHz band. In all three panels, data are binned into 1.5$\arcdeg$ independent longitude bins, following \cite{Ordog_2019}. For all diffuse emission data bins the error bars are one standard deviation divided by the square root of the average number of beams contributing data to the bin. For the CGPS point source RMs the error bars are the standard deviation divided by the square root of the number of sources in each bin. 
 }
	\label{fig:RMversuslong}
\end{figure*}

In \cite{Ordog_2019} we noted the patch of strongly negative RMs in the CGPS ST-only RM map located at the southern edge ($171\arcdeg<\ell<176\arcdeg$, $b<-2\arcdeg$) of the ``bow tie'' H\textsc{ii} complex. This is a large H\textsc{ii} complex, covering approximately $170\arcdeg<\ell<176\arcdeg$, $-3\arcdeg <b<4\arcdeg$ (Figure~\ref{fig:fullmap6}) and shaped like a sideways bow tie, described in \cite{Gao_2010}. In \cite{Ordog_2019} we highlighted that this negative patch led to the dip in RMs observed as a function of Galactic longitude (Figure~\ref{fig:RMversuslong}a). The presence of negative RMs in this region is strongly enhanced by the inclusion of the single-antenna data (comparing Figure~8 in \cite{Ordog_2019} to Figure~\ref{fig:fullmap6} in this paper). In particular, it becomes evident that the negative RM patch at the southern edge is isolated from a positive RM region located to the north of it (centered on $ b\approx-1.5\arcdeg$), but that there is another negative RM region in the northernmost half of the ``bow tie'' ($b>0\arcdeg$). The sensitivity to these large-scale RM structures is contributed by the single-antenna data (Figure~\ref{fig:bowtie}d), but in the small region where the aperture-synthesis-only data satisfy the S/N threshold, their RMs agree with their combined full-scale counterpart, as seen in Figure~\ref{fig:bowtie}e,f and in the example linear fit in Figure~\ref{fig:linear_fits}c.

\cite{Gao_2010} note a region of bright polarized emission at 5~GHz near the southern end of the ``bow tie'' ($\ell\approx172.5\arcdeg$, $b\approx-3\arcdeg$), which they attribute to a Faraday screen enhancing the observed PI by rotating the polarization angle of the background emission to add constructively to the foreground emission. While we do not see a matching PI enhancement at 1.4~GHz, this region does approximately correspond to the strongly negative RMs we observe. It is possible that the foreground and background add constructively at 5~GHz, but not at 1.4~GHz. It will be informative to investigate this region further in other frequency ranges, such as the lower frequencies of GMIMS-LBN with DRAGONS and CHIME.

In the $\ell=173\arcdeg$ H II complex, the inclusion of the single-antenna data allows for much larger RM structures to be observed in comparison with the aperture-synthesis data alone, due to the spatial smoothness of the polarized emission over a large part of the region. With the sensitivity to the large-scale structure provided by the single-antenna data, the aperture-synthesis data then reveal the small-scale perturbations present in the RM values.

\subsection{Depolarization and Bandwidth Effects}\label{sec:discussion_polhorizon}

Across most of the CGPS region at $\ell > 92\arcdeg$ the combined full-scale RM map consists of large, smooth structures that are mostly traced by the single-antenna component alone. Calculating RMs from the GMIMS-HBN data alone (using the CGPS bands) produces a map largely similar to the full-scale RM map. The similarity between the Galactic longitude trends traced by the two is shown in Figure~\ref{fig:RMversuslong}b. However, this pattern differs in a number of ways from the longitude dependence of the aperture-synthesis-only diffuse emission and compact-source RMs \citep[Figure~\ref{fig:RMversuslong}a this paper; Figure~4 in][]{Ordog_2019} and the longitude dependence of the GMIMS-HBN peak Faraday depths (Figure~\ref{fig:RMversuslong}c).

The systematic change from positive to negative RM values near $\ell=60\arcdeg$, followed by a gradual rise back toward RM$=0\radmsq$ near $\ell=180\arcdeg$ traced by the aperture-synthesis-only data (Figure~\ref{fig:RMversuslong}a) does not appear at all in the single-antenna-only or the combined full-scale RMs (Figure~\ref{fig:RMversuslong}b). For the case of the single-antenna-only data, this discrepancy may at least in part be due to beam depolarization and the polarization horizon effect \citep{Uyaniker_2003}, which can limit the depths probed by any particular instrument, due to signals from beyond a maximum distance being completely depolarized. The larger beam of the GMIMS-HBN data may be limiting the depths probed to more nearby regions in the single-antenna-only data than in the DRAO ST aperture-synthesis data, thereby tracing different magnetic field structures. 

The cause of the discrepancy between the large scales traced by the combined full-scale RMs and the aperture-synthesis-only data is less clear, as the combined data should in principle produce the same results as a 600~m single-antenna telescope, which would not be subject to the same depolarization effects as the 26~m single-antenna telescope. Although much of the map is masked out owing to low S/N in the aperture-synthesis-only data \citep[as seen in Figures~\ref{fig:sh216},~\ref{fig:ic443},~\ref{fig:bowtie}, and described in][]{Ordog_2019}, those regions still contribute to the full-scale dataset. However, the RM patterns appear to remain dominated by the 26~m component in the combined data. At least in the case of this study, polarized emission may be dominated by the large scales of the single-antenna data so that any emission from beyond the single-antenna polarization horizon at the small scales of the aperture-synthesis data is subdominant in the combined maps. It may be necessary and informative to examine the RM maps produced from the single-antenna and aperture-synthesis components of any combined dataset separately in addition to the full-scale map. Broadband polarization surveys in the same frequency range as the present study, with larger single-antenna telescopes (such as PEGASUS), may help shed light on this question.

We also compare the peak Faraday depth as a function of Galactic longitude for the full GMIMS-HBN 1280-1750~MHz dataset in the CGPS region by determining a fitted peak to each LOS in the Faraday depth cube and averaging the values in the same longitude bins as the RM data. The result, shown in Figure~\ref{fig:RMversuslong}c, traces a similar pattern to the GMIMS-HBN single-antenna-only and the combined full-scale RMs (Figure~\ref{fig:RMversuslong}b) but with much smaller magnitudes of Faraday depth (note the difference in scales shown in panels b and c of Figure~\ref{fig:RMversuslong}). This is likely the effect of an inadequate `lever-arm' in the linear fits to polarization angle versus $\lambda^2$. For most lines of sight we do not expect the magnetoionic medium to be perfectly Faraday simple, and small fluctuations can occur within the 35~MHz bandwidth used for the RM calculations that lead to generally larger RMs than if a broader frequency range were used. Examples of this are shown in the bottom panels of Figure~\ref{fig:linear_fits}, where the linear fits in the 35~MHz range do not capture the overall trend in polarization angle versus $\lambda^2$. Even in the case of relatively simple deviations from a Burn slab scenario \citep[][uniform electron density and magnetic field]{Burn_1966}, an RM calculated over a relatively narrow $\lambda^2$ range can be highly dependent on the exact $\lambda^2$ range used, as illustrated in Figure~7 of \cite{Ordog_2019}. Consequently, while trends in diffuse emission RM patterns can be traced using narrowband data, the specific values of the RMs should be interpreted with caution.

\section{Conclusions}\label{sec:conclusion}
For the first time, we have carried out an experiment combining single-antenna and aperture-synthesis polarization data across multiple frequency channels. Although the number of frequency channels is small (only four), the resulting dataset should, in principle, fully describe the RMs of the diffuse emission and reveal the full range of spatial scales present in the Galactic Faraday depth sky, down to arcminute resolution. This study is limited to a narrow bandwidth of 35 MHz imposed by the aperture-synthesis contribution from the DRAO ST and the extent in $uv$-plane coverage provided by the GMIMS-HBN component, which is not quite sufficient to completely fill in low spatial frequencies missed by the DRAO ST sampling. Despite these limitations, we have demonstrated that the power of aperture synthesis is vastly expanded when single-antenna data are incorporated into polarization images, specifically for calculating diffuse emission RMs. This has not been previously attempted for large areas of the sky such as the Galactic plane. This experiment has revealed a number of interesting polarization structures that differ in appearance in the absence of either the single-antenna or the aperture-synthesis component. It has also highlighted differences in large-scale Faraday depth patterns that are traced by single-antenna compared to aperture-synthesis data, as well as broadband compared to narrowband data.

Future studies of the magnetoionic medium will require arcminute resolution of the diffuse polarized emission to probe the detailed role of magnetic fields in interstellar processes on physical scales closer to the scales at which individual stars impact their surroundings. The ideal approach to these studies will be to combine aperture-synthesis and large single-antenna polarization data across a wide frequency range ($\sim$ 300-1800~MHz). This will ensure full spatial scale coverage down to the resolution of the aperture-synthesis component, while allowing for multiple Faraday rotation components along the LOS to be resolved, as well as broadened Faraday rotation features, an expected signature of mixed emission and rotation. The upgraded DRAO ST, with simultaneous coverage of 400-800~MHz and 900-1800~MHz in 13 kHz channels, will provide the aperture-synthesis component to current and future Northern Hemisphere single-antenna GMIMS surveys. Our experiment in combining the existing DRAO ST polarization data across its 35~MHz bandwidth with data from GMIMS-HBN paves the way for broadband application of the method, which will yield enhanced resolution and sensitivity to broad structures in both the spatial and Faraday depth domains.

\section*{Acknowledgments}
This paper relies on observations obtained using telescopes at the Dominion Radio Astrophysical Observatory, which is located on the traditional, ancestral, and unceded territory of the syilx people. We benefit enormously from the stewardship of the land by the syilx Okanagan Nation and the radio frequency interference environment protection work by the syilx Okanagan Nation and DRAO. We acknowledge DRAO staff for their work on the site and the telescopes used in this work. DRAO is a national facility operated by the National Research Council Canada.

A.O. is partly supported by the Dunlap Institute at the University of Toronto. This research has been supported by Discovery grants from the Natural Sciences and Engineering Research Council to J.C.B., T.L.L., and A.S.H. A.B. acknowledges financial support from the INAF initiative ``IAF Astronomy Fellowships in Italy'', grant name MEGASKAT. M.H. acknowledges funding from the European Research Council (ERC) under the European Union's Horizon 2020 research and innovation programme (grant agreement No 772663). 

This work benefited from discussions during the program ``Towards a Comprehensive Model of the Galactic Magnetic Field'' at Nordita in April 2023, which is partly supported by NordForsk. In processing, analyzing, and presenting our data we have used the excellent DRAO Export Package developed by our late colleague Lloyd Higgs (1937-2020). We thank the anonymous referee for a thorough and thoughtful report, which has led to improvements to the manuscript.

%

\vspace{5mm}
\facilities{DRAO:26m, DRAO:Synthesis Telescope}
\software{DRAO Export Package \citep{Higgs_1997}; astropy \citep{Astropy_2022}; matplotlib \citep{Hunter_2007}; numpy; RM-tools \citep{Purcell_2020}.}

\section*{Data Availability}
FITS files containing the combined DRAO ST and DRAO 26 m polarization data products (RM, PI, Stokes $Q$ and $U$) are
available through the Canadian Advanced Network for Astronomical Research (CANFAR), doi: \href{https://doi.org/10.11570/25.0003}{10.11570/25.0003}.







\bibliography{refs}{}
\bibliographystyle{aasjournal}



\end{document}